\documentclass[preprint2]{aastex}
\usepackage{natbib}

\newcommand{\Ha}{\mbox{H$\alpha$}}      
\newcommand{\HII}{\ion{H}{2}}    
\newcommand{\NII}{\ion{N}{2}}     
\newcommand{\dlam}{\mbox{$\lambda\lambda$}}

\newcommand{\perpix}{pixel$^{-1}$}

\newcommand{\Teff}{\mbox{$T_{\rm eff}$}}

\newcommand{\Msun}{\mbox{$\cal M_{\odot}$}}

\newcommand{\MV}{\mbox{$M_V$}}
\newcommand{\Mv}{\mbox{$M_V$}}

\newcommand{\RV}{\mbox{$R_V$}}
\newcommand{\Rv}{\mbox{$R_V$}}

\newcommand{\leff}{\mbox{$\lambda_{\rm eff}$}}

\newcommand{\persec}{s$^{-1}$}
\newcommand{\percmsq}{cm$^{-2}$}

\newcommand{\as}{\mbox{$^{\prime\prime}$}}

\newcommand{\fuvmag}{\mbox{\it FUV~}}
\newcommand{\nuvmag}{\mbox{\it NUV~}}
\newcommand{\ebv}{\mbox{$E(B-V)$}}
\shortauthors{Efremova et al.}

\shorttitle{The Recent Star Formation in NGC\,6822: a UV Study}

\begin{document}

\title{The Recent Star Formation in NGC\,6822: an Ultraviolet Study}
\author{
Boryana V.Efremova\altaffilmark{1}, Luciana Bianchi\altaffilmark{1}, David A.Thilker\altaffilmark{1}, 
James D.Neill\altaffilmark{2}, Denis Burgarella\altaffilmark{3}, 
Ted K.Wyder\altaffilmark{2}, Barry F.Madore\altaffilmark{4}, Soo-Chang Rey\altaffilmark{5}, 
Tom A.Barlow\altaffilmark{2},Tim Conrow\altaffilmark{2}, Karl Forster\altaffilmark{2}, Peter G.Friedman\altaffilmark{2}, 
D.Christopher Martin\altaffilmark{2}, Patrick Morrissey\altaffilmark{2}, Susan G.Neff\altaffilmark{6},  
David Schiminovich\altaffilmark{7}, 
Mark Seibert\altaffilmark{4}, Todd Small\altaffilmark{2}}
\altaffiltext{1}{
Department of Phys.\& Astron., Johns Hopkins
University, 3400 N.Charles St., Baltimore, MD\,21218
(boryana,bianchi@pha.jhu.edu)}
\altaffiltext{2}{California Institute of Technology, 1200 E. California Blvd., Pasadena, CA 91125, USA}
\altaffiltext{3}{Laboratoire d'Astrophysique de Marseille, BP 8, Traverse du Siphon, 13376 Marseille Cedex 12, France}

\altaffiltext{4}{Observatories of the Carnegie Institution of Washington, 813 Santa Barbara Street, Pasadena, CA 91101}
\altaffiltext{5}{"Department of Astronomy and Space Science, Chungnam National University, Daejeon 305-764, Republic of Korea}
\altaffiltext{6}{Laboratory for Astronomy and Solar Physics, NASA Goddard Space Flight Center, Greenbelt, MD 20771, USA}
\altaffiltext{7}{Department of Astronomy, Columbia University, New York, NY 10027}

\begin{abstract}

We characterize the star formation in the 
low metallicity galaxy NGC\,6822 over the past few hundred million years,
using $GALEX$ far-UV (FUV, $1344-1786$\,\AA) and near-UV (NUV, $1771-2831$\,\AA) imaging, and 
ground-based \Ha\ imaging. 
From $GALEX$  FUV image, we define 77 star-forming (SF) regions 
with area $>860$\,pc$^2$, and surface brightness $\lesssim$26.8 mag\,(AB) arcsec$^{-2}$, within
0.2$\deg$ (1.7kpc) of the center of the galaxy. We estimate the extinction by interstellar dust in each 
SF region from  resolved photometry of the hot stars it contains: \ebv\ ranges  
 from the minimum foreground value of 0.22~mag up to 0.66$\pm$0.21~mag.  
The integrated FUV and NUV photometry, compared 
with stellar population models, yields
 ages of the SF complexes up to a few hundred Myr, and 
masses from $2\times10^2$\,\Msun\ to $1.5\times10^6$\,\Msun.
The derived ages and masses strongly depend on the assumed type of interstellar selective extinction, which
we find to vary across the galaxy.
The total mass of the FUV-defined SF regions translates into an 
average star formation rate (SFR) of  
$1.4\times10^{-2}$\,\Msun\,yr$^{-1}$ 
over the past 100 Myr, 
and SFR\,=$1.0\times10^{-2}$\,\Msun\,yr$^{-1}$ 
in the most recent 10\,Myr.
The latter is in agreement with the value  that we derive
from the \Ha\ luminosity, SFR\,=0.008
\Msun\,yr$^{-1}$.  The SFR in the most recent epoch  becomes higher if we add 
the SFR=0.02\Msun\,yr$^{-1}$ inferred from far-IR measurements, which trace  star formation
still embedded in dust (age $\lesssim$ a few Myr).  

\end{abstract}



\section{Introduction}
Continuum fluxes in the ultraviolet (UV) and infrared (IR) spectral regions, 
and \Ha\ line emission, are the main indicators 
of star-formation activity in distant galaxies
(see e.g. Kennicutt 1998).  
The UV flux is a direct tracer of young massive stars, whose energy is mostly emitted in this spectral region, 
 \Ha\ emission originates from interstellar gas ionized by the most massive stars, 
and the far-IR emission is produced by dust particles re-emitting reprocessed UV stellar light.  

Integrated measurements of these fluxes can be translated into 
star-formation rates of galaxies, but 
additional information is needed.
First,  observed fluxes need to be corrected for extinction by interstellar dust, both foreground
(by Milky Way (MW) dust along the line of sight), and internal (within the galaxy).
Reddening is particularly significant at UV wavelengths (see e.g. Bianchi 2011). 
Stellar population models with adequate star-formation history (SFH) are then used to transform the continuum and line-emission
luminosities into SFRs. 

The UV photometry of star-forming galaxies is usually corrected for interstellar extinction 
assuming  a MW-type selective extinction with $\Rv=3.1$ 
\citep{Cardelli89} for the foreground component,
and the \citet{Calzetti01} extinction curve for internal extinction.
The amount of extinction is sometimes estimated by comparison of UV and far-IR fluxes
\citep{Calzetti05,Cortese06, Boissier07,Meurer09}.
Such method assumes that the intrinsic FUV-NUV color is known, however its value is strongly
varying with age for young starbursts (e.g. Bianchi 2009, 2011), and that UV and far-IR fluxes
are emitted by the same population, which is often not the case. 

In unresolved distant galaxies only integrated measurements are possible, and a global extinction 
correction and  star-formation history (SFH) must be assumed for interpreting such
measurements. 
On the other hand,  in nearby galaxies  individual 
SF regions can be measured, and their stellar content studied in detail 
(e.g. Bianchi \& Efremova 2006, Bianchi et al. 2011, 2010, 2001, Kang et al. 2009, and references
therein). Therefore, the dust properties can be explored in a variety of local environments,
providing information on  the interplay of dust and star formation, and a calibration
of star-formation indicators in distant galaxies. 

Deep imaging   in FUV and NUV for hundreds of nearby galaxies were obtained with 
the {\it Galaxy Evolution Explorer} ($GALEX$) \citep{Martin05, Morrissey07} 
as part of the  Nearby Galaxy Survey (NGS)  \citep{Bianchi03a, Bianchi03b, Bianchi09, dePaz07}.
The wide-field UV imaging provides a characterization of the young stellar populations across 
the whole  extent of these galaxies, and 
can be used,  with complementary optical data,  to infer their star-formation history and SFR. 

In this paper we perform a comprehensive study  of the young stellar populations
in the Local Group low-metallicity galaxy NGC\,6822,  the nearest 
SF galaxy currently with no massive neighbor.
We identify and define SF regions from  $GALEX$ wide-field imaging in FUV,
where the hottest, youngest stars are more prominent, throughout the extent of the galaxy. 
We use  integrated photometry of these regions in FUV and NUV, and  
complementary  \Ha\ emission-line imaging, as well as information from
 resolved stellar photometry, to investigate the star-formation in this galaxy 
during the past $\sim100$\,Myr, and the characteristics of interstellar extinction. 
The study of this galaxy, together with results for Local Group galaxies of
other types, contributes one piece to a broader puzzle, aimed at 
 understanding the modalities of  star formation in  differing environments, and the role of dust.

This benchmark galaxy was chosen to complement the study 
 by Kang et al. (2009) of M31, and of other Local Group galaxies by \citet{Bianchi10,Bianchi10b},
because of its low metallicity and vicinity ($494$\,kpc, McAlary et al. 1983) 
and the abundant information available from resolved stellar population studies  
 with $HST$ multi-band  imaging (Bianchi et al. 2001, Bianchi \& Efremova 2006), 
CTIO $UBV$ imaging (Massey et al. 2007a), 
$VLT$ $UBV$ imaging and extensive spectroscopy (B. Efremova et al. 2011, in preparation). 
NGC\,6822's metallicity is believed to be subsolar: measurements by \citet{Muschielok99} of
three B--type supergiants, and by \citet{Venn01} of two A--type supergiants both yield $Z\approx0.006$.

The paper is arranged as follows. 
In Section~\ref{sec:obs} we define SF regions from $GALEX$ FUV imaging, and measure 
their integrated fluxes in FUV and NUV;
we also use the CTIO \Ha\  imaging of \citet{Massey07b} to define 
and measure regions of \Ha\ emission. In Section~\ref{sec:analysis} the 
integrated measurements of the SF regions are analyzed with stellar population models 
to derive their ages and masses, after the 
interstellar extinction is estimated from the massive stars within each SF region. 
The results are discussed in Section~\ref{sec:results}, and summarized in Section \ref{sec:summary}.

\section{Observations.  Detection and  Photometry of the Star-Forming Regions}
 \label{sec:obs}
\subsection{UV imaging}
\label{sec:UVdata}

We used $GALEX$ images in FUV ($\leff=1539$\,\AA, FWHM\,$\approx270$\,\AA),
and NUV ($\leff=2316$\,\AA, FWHM\,$\approx615$\,\AA) with resolution 
4\as.2 (FUV) and 5\as.3 (NUV) \citep{Morrissey07},
corresponding to $\sim 12$\,pc at the distance of NGC\,6822.
The images are sampled with  1\as.5 pixels. 

The $GALEX$ images of NGC\,6822 were taken on Aug 20th 2005 as 
part of the NGS program, with
exposure times of 4654 sec (FUV) and 6198 sec (NUV). The data was downloaded  from the MAST archive.
The 1.2 degree diameter $GALEX$ field of view  is centered at RA\,$=19\ 44\ 57.37$, Dec\,$=-14\ 47\ 33.32$, 
near the center of the galaxy (RA\,$=19\ 44\ 57.8$, Dec\,$=-14\ 48\ 11$, FK5 2000). 
NGC\,6822, with an optical diameter (at $\sim25$ mag/arcsec$^2$) of 15\arcmin.6  
\citep{Karachentsev04},  is contained in the central portion of the $GALEX$ image, which is 
shown in Fig.~\ref{fig:fuv_ha}.

\subsection{UV Source Detection}
\label{sec:sourcedet}
We identified the SF regions using the $GALEX$ FUV image, which unambiguously reveals 
the young, hot massive stars not heavily embedded in interstellar dust.
We followed the general method of Kang et al. (2009), adapted to the case of NGC\,6822.
We  defined contours of regions with FUV
surface brightness $\geq$ $3\sigma$ above the background.
An important issue in defining extended source contours and measuring their flux
is the background estimate.
Several approaches were used to find the best method for background evaluation (see also the discussion in
Kang et al. 2009).
For the purpose of source detection only, we constructed a background image applying
a two-step  circular median filter (64 pixels diameter, $\approx$1.5arcmin)  to the FUV intensity map (``int'' file).
The first pass of the filter identifies pixels which belong to 
localized peaks via masking pixels brighter than the local 
median background estimate. The second pass 
of the filter operates only on the final list of 
non-peak  pixels to obtain a background image less biased by 
substructure than a one-pass median filter. 
The diameter of the median filter was chosen to provide a background 
image where measurements of the background for individual sources are closest to 
the median  flux density of the intensity map  images
in 6-pixels wide annuli around the sources, from here on ``local background''. 
The background image produced using the adopted median filter gives sky estimates
slightly lower (by about 0.17/0.18 mag arcsec$^{-2}$ for FUV/NUV)
than the local background. 
For comparison, the background image provided by the pipeline
gives a background estimate lower than the local background  
by 0.52/0.60 mag arcsec$^{-2}$  on average  for FUV/NUV.
A background-subtracted image was constructed, subtracting our background image 
from the intensity map image, and used for source detection. 

Source contours were defined to enclose contiguous pixels with FUV
flux more than $3\,\sigma$ above the background image. This threshold corresponds to an
FUV surface brightness of
0.0015 counts \persec\ \perpix\ (26.8 AB\,mag\,arcsec$^{-2}$) on the background-subtracted image, or average 
0.0025 counts \persec\ \perpix\ (26.2 AB\,mag\,arcsec$^{-2}$) on the intensity map image.

\begin{figure*}
\hskip-1cm
\vskip -4 cm
\epsscale{1.8}
\plotone{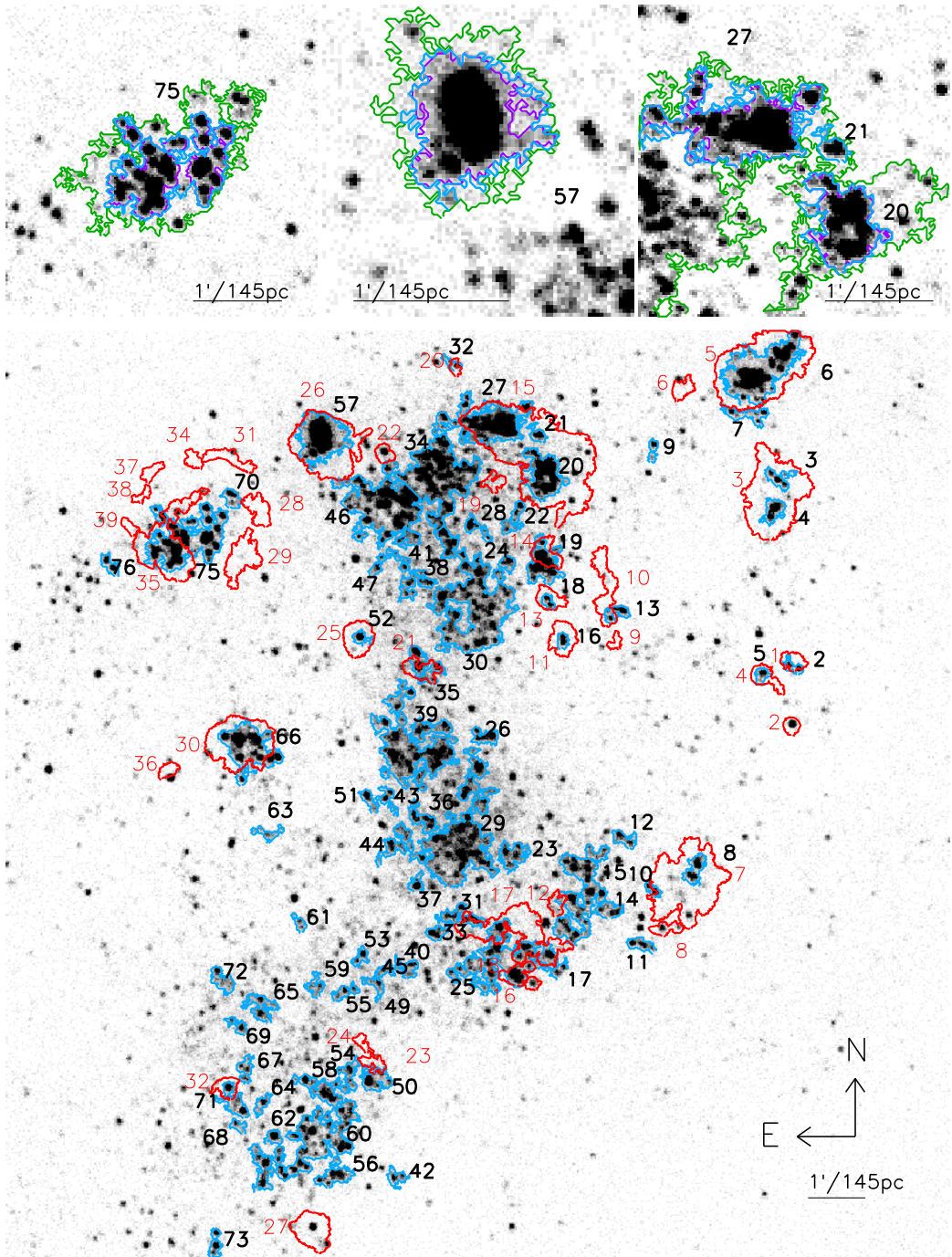}
\vskip -0.5 cm
\caption{
Contours of SF regions (blue) over the $GALEX$ FUV image,
defined with FUV surface brightness $>$ $3\,\sigma$ above the 
background level and area $>150$\,arcsec$^2$ ($860$\,pc$^2$). 
\Ha\ contours ($3\,\sigma$) are shown in red.
Enlargements of regions \# 75, 57 and 27 and 20 in the top panels
illustrate the effect of the threshold choice for contour definition 
($1\,\sigma$ in green, $3\,\sigma$ in light blue, $4\,\sigma$ in dark blue).
\label{fig:fuv_ha} }
\end{figure*}

The effect of the threshold choice on the source-contour definition is illustrated 
 in the top panels of Figure~\ref{fig:fuv_ha}, which show 
 SF regions \# 75, 57, 27, and 20 defined for thresholds of $1\sigma$ (green), $3\sigma$ (light blue), 
and $4\sigma$ (dark blue). A low threshold of 1 or $2\sigma$ would cause regions like
\# 27 and 20 (OB8 and OB6, from here on  `OB' designations are from Hodge 1977)  
to merge, and the main body of 
the galaxy to appear as one large region. If a threshold higher than $3\sigma$ is used, sparse associations 
like region \# 75 (OB15) split into several sources or into individual stars (see also Kang et al. 2009 for 
more discussion on the procedure and the choice of parameters).

The sources defined using the $3\sigma$ threshold follow the distribution of blue stars as shown by  resolved 
stellar photometry from $HST$ imaging (Bianchi et al 2001, Bianchi \& Efremova 2006), and ground-based data
(Massey et al 2007a).
To exclude artifacts, in the initial list  we rejected sources with area less than 16~arcsec$^2$ (7 pixels).

We further restricted the analysis sample to sources larger than 150~arcsec$^2$ 
($\approx 860$\,pc$^2$), 
in order to exclude single stars (mostly foreground) and background objects, 
and to examine SF complexes massive enough that stochastic effects will not be significant 
in deriving ages and masses by model analysis (Section \ref{sec:analysis}). 
Stochasticity may affect the comparison of integrated star cluster photometry with
stellar population models, as was first pointed out by \citet{Girardi95}. 
Quantitative assessment of this effect is still a matter of debate, given that more factors 
are relevant in such analysis, including IMF, metallicity, extinction.
For example, \citet{Fatuzzo08} estimate that for clusters with $\gtrsim$ 1000 stars 
 the IMF is sampled well enough  so that their UV flux is close to model 
predictions for integrated populations (but they also point out that the exact limit may vary with IFM).
Their  analysis concerns statistical distributions of bound, spherical, zero-age  
stellar clusters. Such limit corresponds to $\geq$5 stars more massive than 10\,\Msun\ i.e. earlier than 
spectral type B2V (with the parameters adopted by these authors). 
We will return on this point again later. 

\begin{figure}[htb]
\hskip-1cm
\vskip -0.5 cm
\epsscale{0.9}
\plotone{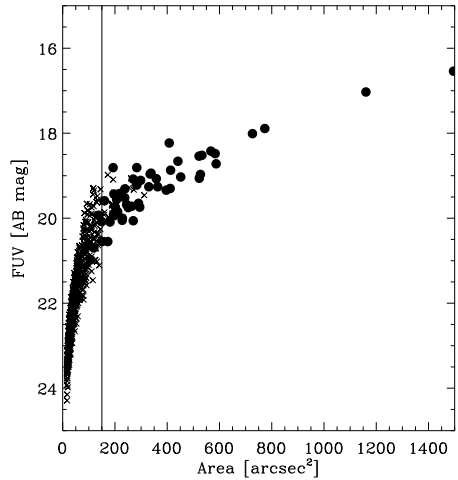}
\vskip -0.5 cm
\caption{FUV magnitude vs. area of FUV-defined SF regions.
Regions containing five or more blue stars (see the text) are marked with
filled circles.  A cut by area at 150 arcsec$^{2}$, adopted for our analysis sample,  is shown with a vertical line. It retains 94\% of the regions containing $\geq$5 blue stars in the sample.
A brightness cut at $FUV\le20.2$\,mag would retain the 
the same fraction of regions with $\geq$5 blue stars, however it would highly increase 
the fraction of regions with less than five stars included in the sample.
 \label{fig:brightcut} }
\end{figure}

We choose to restrict the analysis sample with the area cut of 150~arcsec$^2$
after examining the distribution of stars with
spectral type earlier than B2V\footnote{Selected from the photometry of ~\citet{Massey07a} to have 
 $(B-V)_0<-0.15$ and $\Mv<-2.5$, after the reddening correction is applied, as described in Section 3.1.} 
inside the FUV-defined contours. 
In Fig.~\ref{fig:brightcut} we plot the FUV magnitude
{\it vs} area of the SF regions: those containing $\geq$5 blue massive stars are shown with dots. 
A cut by area of $\ge150$\,arcsec$^{2}$  includes 94\% of these regions
in the sample, and  very few regions containing less than five blue stars (20\%). 
We examined the alternative option of a brightness cut, which is often used in studies of  more 
distant galaxies. Such criterion would either include fewer  SF regions with five or more 
blue stars, or  more  regions with less than five blue stars,  in the analysis sample. 
For example, a brightness cut at $FUV\le20.2$\,mag  includes in the analysis sample
94\% of the regions with $\geq$5 blue stars 
(the same fraction as our area cut of $\geq$150\,arcsec$^{2}$),
but 44\% of the selected regions would contain less than five blue stars.
Therefore, a cut by area better satisfies our requirement of  a minimum number of blue stars
within a source contour, including in the analysis sample as many as possible of the clusters having
$\leq$5 massive stars, and as few as possible clusters with $<$5 massive stars. 
A brightness or luminosity cut would also strongly be affected by extinction, or
extinction correction (see Section \ref{sec:parameters}).
We point out that this criterion may not necessarily be the best choice for more distant
galaxies where similar data would give a lower spatial resolution, or for galaxies where
star formation is less sparse. In the specific case of NGC6822, such  criterion, more precise than a
luminosity cut for our purpose, could be tested and tuned given the vicinity of the galaxy and   
the detailed information on its stellar population. 
Finally, a cut by area may eliminate young compact clusters, and may be undesirable in 
disk galaxies  for example, where young compact star clusters abound 
(e.g. Bianchi et al. 1999, Chandar et al. 1999 for M33;  Hodge et al 2010, and Kang et al in preparation, for M31).
In NGC6822 there are very few such compact clusters and their exclusion would not change our 
results. This work aims at the detection of unbound OB associations and SF complexes, 
not compact  star clusters.
Another advantage of the area cut is that it effectively excludes foreground stars.

The resulting analysis sample includes 77 FUV-defined sources with area $\geq$150~arcsec$^2$
and brightness $\lesssim$ 26.8mag~arcsec$^{-2}$,
within a 0.2\,deg radius (1.72~kpc) of the center of NGC\,6822.
The 0.2\,deg radius is 1.5 times the  optical semi-major axis  of the galaxy (7\arcmin.8
at $\sim25$ mag/arcsec$^2$) given by \citet{Karachentsev04}.  
\Ha\ emission is detected out to a radius of R$_{\Ha}=1.65$\,kpc\ \citep{Hunter04} (see also Section 2.3),
and the HI disk \citep{deBlok06} also exceeds the optical size of the galaxy (see also Bianchi 2011). 
Our sample extends to a slightly larger area than that of \citet{Melena09}\footnote{
\citet{Melena09} sample of SF regions is within 1.65\,kpc, using the coordinates of their
Table 2, in spite of their claim that it extends to 6\,kpc.}.

The areas of the selected SF regions range from 150 to 5400 arcsec$^2$ ($860 - 3\times10^4$ pc$^2$).
Table~\ref{tbl:int_ph} gives identification, coordinates of the ``centroids'' 
(the median $\alpha$\ and $\delta$\ values of the pixels included in the contours), 
and areas of the FUV-defined regions, ordered by increasing R.A.
The contours are shown in blue in Figure~\ref{fig:fuv_ha} over the $GALEX$ FUV image. 
In the next section we describe the photometry measurements, which are used in Section \ref{sec:analysis}
to derive ages and masses.

\subsection{UV Photometry of the star-forming regions}
\label{sec:phot}

For photometric measurements we used the intensity map (``int'') images (in units of counts\ \persec\ \perpix)
generated by the $GALEX$ pipeline 
 dividing the count map by the relative response map\footnote{\it http://galexgi.gsfc.nasa.gov/Documents/ 
ERO\_data\_description\_3.htm}.   
We measured the FUV and NUV flux of each SF region within its FUV-defined contour, and the local background. 
 The  background was measured over an area defined by smoothing
the source contour and expanding it by 3 pixels (inner background contour) and 9 pixels
(outer contour), i.e. creating a 6 pixels wide `annulus' around the source, which follows its shape. 
The median of the 
flux \perpix\ in the background region, excluding portions of nearby sources  falling in the 
background annulus, was then subtracted from every pixel inside the
source contour. The conversion from [counts\, ~\persec] to magnitudes in the AB photometric system 
was performed using  zero-points ZP\,$=18.82$mag (FUV) and $20.08$mag (NUV) \citep{Morrissey07}.
We calculated the photometric errors as 
$\Delta{\rm  mag}\approx 2.5/\ln(10)\times \frac{N}{S}$  = $1.09\frac{N}{S}$,
where $S$ is the flux from the source in the aperture and $N$ is the quadratic sum of all the 
noises affecting the image.
$S$ is expressed by $S=F\times EPADU$, where $F$ is the flux in counts, and $EPADU$ is the 
conversion factor from ADU to e$^{-}$,  for $GALEX$ $EPADU=1$.
We consider the Poisson noise of the photon flux, and the background fluctuations to dominate,
so we used the following expression to estimate the noise:
$N=\sqrt{N_{source}^2 + N_{background}^2 + A\sigma^{2}}=\sqrt{S + S_{background} + A\sigma^{2}}$,
where $A$ is the area of the source  (in pixels) and $\sigma$ is the standard deviation among
the pixels in the background annulus (in counts).
The resulting \fuvmag and \nuvmag magnitudes and their errors are 
listed in Table~\ref{tbl:int_ph}.

The total flux from the selected SF regions is 45\% of the integrated FUV flux from NGC\,6822
($F_{FUV\rm tot}=1.05\pm 0.07\times10^{-9}$ erg s$^{-1}$ cm$^{-2}$ or $FUV_{tot}=12.1\pm0.07$ mag AB) 
and 35\% of the  total NUV flux ($F_{NUV\rm tot}=1.09\pm 0.04\times10^{-9}$ erg s$^{-1}$ cm$^{-2}$ 
or $NUV_{tot}=11.7\pm0.04$ mag AB), measured from the pipeline sky-subtracted image 
in an aperture of 0.2 deg radius.
The flux not included in our SF-regions  comes from smaller sources excluded by our area cut
(10\% of $F_{FUV\rm tot}$), from older diffuse populations, and from scattered emission from SF regions.
The fraction of flux included in the selected SF-sites 
is lower in NUV than in FUV because foreground stars and diffuse light from older populations
are more conspicuous at longer wavelengths.

\subsection{\Ha\ Emission Sources}
\label{sec:haphot}

We also used the publicly available CTIO \Ha\ image from the 
survey of \citet{Massey07b} to define contours of \Ha\ emitting regions.
The \Ha\ image has an exposure of 300 sec, a scale of 0\as.27 \perpix, and 
a resolution of 0\as.9 (2.2 pc at the distance to NGC\,6822).
We used the $V$ and $R$ images from the same survey \citep{Massey07a}
to correct the \Ha\ image for continuum, by subtracting from it a linear combination 
of the $V$ and $R$ images, scaled to match the intensity of the continuum sources.
We define contours of \Ha\ emitting regions using a
threshold of  $3\sigma$ above the background, corresponding
to a surface brightness of $3\times10^{-17}$ erg \persec cm$^{-2}$  arcsec$^{-2}$.
The \Ha-defined contours are drawn in red in  Fig.~\ref{fig:fuv_ha}.
They generally follow the \Ha\ contours defined by \citet{Hodge88,Hodge89}
in a similar way and using a threshold of $2\times10^{-17}$ erg \persec cm$^{-2}$  arcsec$^{-2}$.

We used the calibration factor of 1 count\,\persec =$1.8 \times 10^{-16}$ergs \persec cm$^{-2}$
given by \citet{Massey07b} for emission line sources.
We did not attempt to correct for [\NII] emission line contamination, 
which we expect to not exceed a few percent of the flux.
The average $F_{[NII]\lambda6584}$/$F_{\Ha}$ in the \HII\ regions measured by \citet{Pagel80} is about 6\%;
adopting $F_{[NII]\lambda 6548}$/$F_{[NII]\lambda 6584}$$\approx$\,1/3, we derive 
$F_{[NII](\lambda 6548+\lambda 6584)}$/$F_{\Ha\,}$$\approx$\,8\%.
Since both [\NII]\dlam\,6548, 6584 fall in the wings of the 50\AA\ -wide filter passband centered at \Ha , 
we expect the actual contribution from [\NII] emission to be of the order of $2-3$\%.

The \Ha\ measurements are listed in Table \ref{tbl:int_phha}. In the last column we give
the cross-identifications with \Ha\ sources previously defined by \citet{Hubble25} (H followed by roman number), 
 \citet{Hodge77} (Ho followed by arabic number),  \citet{Kinman79} (K followed by a Greek letter), 
 \citet{Killen82} (KD followed by arabic number), and \citet{Hodge88} (HK followed by arabic number).
For regions with \Ha\ surface brightness $>1\times10^{-16}$ ergs \persec cm$^{-2}$  arcsec$^{-2}$, 
our measurements agree within $\sim25\%$ with those 
by \citet{Hodge89}, 
after de-correcting the latter for extinction with A$_{\Ha}=0.9$ (O'Dell et al 1999). 
The total \Ha\ flux from the \Ha\ emitting regions, $F_{\Ha}=1.8\pm 0.2\times 10^{-11}$\,erg \persec\,\percmsq\ 
(the sum of all measurements in Table \ref{tbl:int_phha}), 
is higher by about 2\% than the total flux from Table 1 of \citet{Hodge89}, most of the difference coming from
faint \HII\ regions not covered by the imaging of \citet{Hodge88}.
We further checked our \Ha\ calibration and continuum-source subtraction by measuring the flux from
 the \HII\ regions Hubble~V and Hubble~X surrounding OB8 and OB13 in 42\as\,$\times 42$\as\ square apertures,
similar to the ones used by  \citet{ODell99}: our measurements agree within 10\% with the values given
by these authors.

\section{Analysis}
\label{sec:analysis}

\begin{figure*}[htb]
\hskip -1.cm
\epsscale{1.5}
\plotone{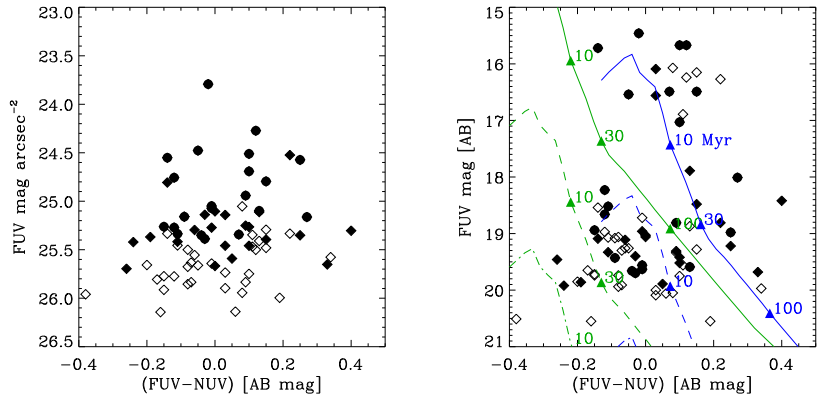}
\caption{
UV color-magnitude diagram of the SF regions 
(left: FUV surface brightness, right: FUV magnitudes).
The sources with associated \Ha\ emission are marked with dots, the others
with diamonds (filled if the background is $<$30\% of the total flux within the source contour).
The right panel shows also synthetic SSP  model magnitudes 
at different ages, scaled to cluster masses of $1\times10^4$/$1\times10^3$/$1\times10^2$\,\Msun\ 
(solid/dashed/dash-dotted lines).
Models are reddened with MW-type exinction for a foreground reddening of $\ebv=0.22$ (green lines), 
and with an additional  $\ebv=0.15$ using LMC2-type extinction  (blue lines).
Three age values (10, 30, and  100\,Myr) are marked with triangles and labelled.
}
\label{fig:galcol} 
\end{figure*}

The UV color-magnitude diagram  of the FUV-selected SF regions 
is shown in Figure~\ref{fig:galcol}: in the left panel we plotted 
the FUV surface brightness (mag arcsec$^{-2}$),
and in the right panel the total FUV magnitude. 
In the right panel we also show FUV synthetic magnitudes for single-burst 
(coeval) populations (Bianchi 2011) at various ages, scaled for total stellar  
masses of $1\times10^4$\,/$1\times10^3$/$1\times10^2$ \Msun\ (solid/dashed/dash-dotted lines).
Models are reddened with $\ebv=0.22$ mag (the assumed foreground extinction) using MW-type extinction (green lines)
and with an additional $\ebv=0.15$ using the extinction curve derived for stars in the LMC2 
star-forming region by \citet{Misselt99} (blue lines). 
For SF regions located in the main body of the galaxy, the local background 
(which includes the diffuse older populations, more conspicuous in the NUV band)
is significant, and therefore there is always a concern that 
its subtraction may lead to greater uncertainty than the formal errors indicate.
An error in FUV-NUV color would propagate to an error in the derived age, and consequently 
on the derived mass. 
To verify that no bias is introduced by high background subtractions, we plotted with empty diamonds 
the sources for which the background amounts to more than 30\% of the flux in the source contour. 
These high-background sources are distributed in color-magnitude space not differently from the other sources,
confirming that no biases have been  introduced by the critical background estimate procedure.

In the following sections we estimate  ages and masses of the SF regions by comparing their 
UV photometry  with Simple (single-burst) Stellar Population  (SSP) model colors of different
metallicities (e.g. Bianchi  2007, 2009, 2011),
after the \ebv\ is estimated for each SF region from resolved stellar photometry.
The models are progressively reddened with various types of interstellar dust extinction. 

\subsection{Interstellar Reddening}
\label{sec:ebv}

Flux at UV wavelengths is very sensitive to extinction by interstellar dust, and in order
to derive age and mass of the SF regions from model analysis, we first estimated the 
amount of reddening in each.

We used information from resolved stellar photometry, 
and estimated  \ebv\ of the hot massive stars
(selected with $(B-V)_0<-0.2$ and $\Mv<-2.5$, i.e. earlier than about $\sim$\,B2V) within each contour.
 For several OB associations, \ebv\ values are available for individual stars, derived by 
Bianchi et al (2001)  and Bianchi \& Efremova (2006) from $HST$ multi-band photometry (from UV to optical).
For the most massive stars in six OB associations, we also have $VLT$ spectroscopy 
(B. Efremova et al. 2011, in preparation), which confirms the results from $HST$ photometry. 
For the regions without $HST$ photometry, we  derived \ebv\ using 
the $CTIO$ $UBV$  photometry of \citet{Massey07a},
with the standard ``Q-method'' (e.g. Bianchi \& Efremova 2006, Kang et al. 2009), and  
by comparing the observed ($U-B$),($B-V$) colors with progressively reddened  stellar model colors.

We accounted for extinction in each SF region  using the median \ebv\ value
 of the massive stars it contains.
The values range from $\ebv = 0.22$ (purely foreground extinction) to 0.66\,mag, with a mean of $\ebv =0.36$\,mag, 
and are given in Table~\ref{tbl:int_ph}. The mean \ebv\ values 
are similarly distributed, ranging from $\ebv =0.21$ to 0.51\,mag, with an average of $\ebv = 0.37$\,mag. 
The typical $1\sigma$ scatter (also given in Table~\ref{tbl:int_ph}) around the mean \ebv\ in individual
SF regions is 0.13\,mag.
The wide range of extinction values in the SF regions (not uncommon in star-forming galaxies) 
underscores the importance of accurate
extinction corrections, particularly relevant in the UV regime, for deriving ages and masses. 
Other works adopt a generic assumption, for example 
\citet{Melena09} used a constant extinction of 0.27 mags for all their sample regions in NGC6822,
corresponding to $\ebv =0.05$\,mag of internal extinction in addition to the $\ebv = 0.22$\,mag foreground reddening.
Our results derived for individual  regions (Table \ref{tbl:int_ph}) 
indicate that a higher value is more typical. The model magnitudes plotted in Fig.3 
illustrate how such assumptions affect the derived ages and masses; more model plots showing  the
effects of extinction can be found in Bianchi (2011). 

We assume a foreground extinction of
$\ebv=0.22$\,mag, consistent with the minimum \ebv\ estimated in this work, and with previous 
estimates by Bianchi et al (2001), Bianchi \& Efremova (2006), and \citet{Massey07a}.
 Any additional extinction is considered to originate within NGC6822. 
 
While the derived \ebv\ values are mostly based on optical photometry of the stars,
and do not depend significantly on the type of dust, the selective extinction
A$_{\lambda}$/\ebv\ at UV wavelengths is known to strongly vary with environment.
Therefore, the correction of UV magnitudes, and the resulting ages and masses, 
strongly depend on the adopted   extinction curve (e.g. Bianchi 2011).
In the analysis that follows, we consider four different types of 
 internal extinction, found in the MW and in known  low-metallicity star-forming environments:
 1)  MW-type extinction with $\Rv=3.1$. In this case the  $GALEX$ (FUV-NUV) color
is basically reddening-free (Bianchi 2011 and references therein); 2) the average extinction curve
 derived by \citet{Misselt99} for LMC stars outside the 30 Doradus region (from here on, ``LMC''),
which gives an average color excess ratio for hot stars
($\Teff>10000$\,K) of $E(FUV-NUV)/\ebv\approx1$; 3) the extinction curve
in the LMC 30 Doradus region (from here on ``LMC2'') derived by \citet{Misselt99}, 
yielding $E(FUV-NUV)/\ebv\approx 2$;
and 4) the extremely UV-steep extinction curve derived by \citet{Gordon98}
for SMC stars (from here on, ``SMC''), which yields $E(FUV-NUV)/\ebv\approx5$.

\subsection{Ages and Masses of Star-Forming Regions}
\label{sec:parameters}

We derive ages of the SF regions from their integrated FUV-NUV colors compared with SSP
models for low metallicity populations (see below), and masses from the age and 
UV magnitudes, accounting for reddening. 
We compared results obtained by 
adopting the different extinction curves mentioned in the previous section with 
information from resolved  stellar photometry and \Ha\,, in order to assess what type
of selective extinction is more appropriate. 

We found that a uniform extinction type  is not adequate for the whole sample of SF regions in NGC\,6822.
If we assume ``average MW'' extinction with $\Rv=3.1$  for the whole sample 
(as adopted e.g. by Hunter et al. 2010), the measured colors imply  
ages too old for several regions which show \Ha\ emission and which appear to be  a few Myr old in 
 H-R diagrams from $HST$ photometry (Bianchi et al 2001).
On the other hand, if we use a UV-steep extinction curve, ``LMC2'' for example,  to deredden all
SF regions,  the (FUV-NUV) color for part of the sample is over-corrected, 
such that it appears unrealistic 
when compared with model predictions at any age.
We found that different extinction curves are needed to
bring the ages from $GALEX$ integrated measurements in agreement with 
results from resolved studies for a subsample of well studied regions.
Regions \# 27, 57, 75, 19, 20 (approximately corresponding to OB8, OB13, OB15, OB7, OB6 as 
defined by Hodge 1977), and \# 52,
are included in the $HST$ photometric studies  by Bianchi et al. (2001) 
and Bianchi \& Efremova (2006); $VLT$ spectroscopy of the most massive stars 
confirms the ages derived from $HST$ photometry.
 By comparing results from integrated measurements with  resolved stellar photometry of 
these well studied SF regions (outside the central part of the galaxy, where measurements
are not complicated by significant diffuse older populations),
and with information from \Ha\ emission,  we
derived a criterion for choosing the type of extinction curve, and apply it to the rest of the sample. 
Specifically, the age information for the best studied SF regions, with spectroscopy available for the hottest stars,
 was based on the presence (or absence) of O-type stars, W-R type stars, or B supergiants, as well as on the 
photometric H-R diagram of their stellar population, and
we ruled out extinction curves giving very discrepant results from the $GALEX$ color. 
 For the FUV-bright  regions clearly associated with 
\Ha\ emission, we ruled out extinction curves yielding
ages significantly older than 10Myr from  integrated UV photometry. Finally,  UV-steep
extinction curves were excluded in the cases where they would yield an intrinsic FUV-NUV color bluer than 
any stellar population model at any age.
While  derivation of the actual extinction curve would require UV spectroscopy of several stars in each region, and it
is not possible with broad-band photometry, the representative known curves examined  
give sufficiently different results (Table 3) that some of these assumptions could definitely be excluded in many cases.  
Some consistent trends within the subsample of SF regions with information on their stellar content
allowed us to define general criteria.  

\begin{figure}[htb]
\hskip -1.cm
\epsscale{.85}
\plotone{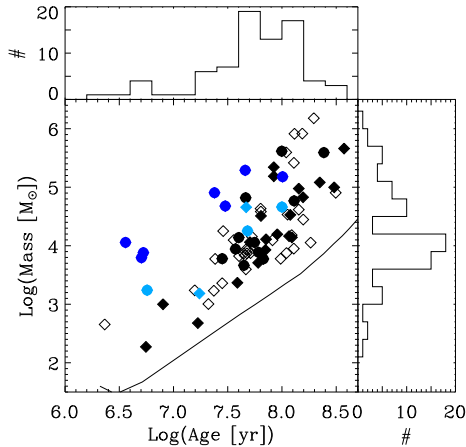}
\caption{
Masses {\it vs} ages  of the FUV-defined SF regions, derived 
from integrated UV magnitudes using SSP models with  metallicity $Z=0.004$. 
We accounted for interstellar extinction using \ebv\ values estimated for each SF region,
and  assuming  a foreground component with $E(B-V)=0.22$ and MW-type dust ($\Rv=3.1$),
and different extinction curves for the  additional internal extinction 
as described in the text 
(black if ``MW, $\Rv=3.1$'', light blue if  ``LMC'', dark blue if  ``LMC2'').
As in the previous figure, UV sources associated with  \Ha\ emission are plotted with dots,
and the rest with diamonds (empty symbols if the background is $>$30\% of the source flux).
The black line shows the detection limit for a source with our minimum area and 
only foreground reddening.
}
\label{fig:ages} 
\end{figure}

For the youngest SF regions, a UV-steep extinction curve brings ages from integrated measurements
into agreement with resolved H-R diagram results. These regions are associated with strong \Ha\ emission 
(plotted with circles in Figs.~\ref{fig:galcol} and \ref{fig:ages}) and typically have
high FUV surface brightness, i.e. the SF is intense and the complex is very compact. 
Therefore, we assumed UV-steep extinction curves to correct for internal extinction
of regions with surface brightness higher than $25.0$ mag arcsec$^{-2}$,
and  MW-type extinction for the rest of the sample.
Among the high surface-brightness SF regions, we found ``LMC2''-type  extinction 
to be preferable for sources with $FUV<17.5$\,mag, and the 
less steep ``LMC'' extinction to be  better for sources fainter (in FUV) than this limit. 

We stress that these criteria were derived {\it ad hoc}, to bring the results from integrated 
measurements into agreement with detailed studies of a subsample. 
$HST$ photometry of sample regions for a wider sample of Local Group galaxies
will be used to expand this comparison (Bianchi et al 2010, 2011).   
 However, the results are not surprising: for example, in a study of  SF sites in M51,  \citet{Calzetti05}
found  starburst-like extinction to be applicable only to sites with strongest star formation.

Two metallicity values, $Z=0.004$ and $Z=0.008$, were explored, encompassing the estimated metallicity of young stars
in NGC\,6822, $Z=0.006$ (Muschielok et al. 1999, Venn et al. 2001).
Models with metallicity $Z=0.008$ yield slightly younger ages, and consequently lower masses of the SF regions.
The effect of  metallicity  on the derived age and mass  varies with age and  extinction type, as discussed by  
Kang et al. (2009; see their Fig.12), and Bianchi (2009, in particular Fig.9), Bianchi (2011).

Ages and masses of the SF regions, derived from integrated FUV, NUV photometry 
using models with metallicity $Z=0.004$, 
and assuming three different extinction curves for internal extinction, are given in Table~\ref{tbl:agmass};
 the last column indicates which values are adopted in our analysis. 
These values are plotted in Figure \ref{fig:ages}. 
The gap in the age distribution at $\sim10$\,Myr is due to a slight degeneracy
of the UV color in the range $6-11$\,Myr, due to RSGs emission (Fall et al. 2009).
The uncertainty in the derived ages caused by  this effect is smaller
than the uncertainty introduced by the scatter in $E(B-V)$, and it does not 
affect our overall results significantly. 
We stress that the masses of the individual SF regions should not be interpreted in terms
of cluster mass function, for two reasons. First, we defined irregular, mostly unbound, complexes. 
Second, we used a constant threshold throughout the galaxy, to derive source contours,   
in the interest of adhering to an objective criterion and a consistent flux limit. This choice 
inevitably may cause some sparse regions to break into subcomponents (but their ages,
and the masses of each, would not be affected), and more diffuse regions to merge.  


The magnitude limit of our sample, from the 3$\sigma$ detection threshold of 26.8 mag\, arcsec$^{-2}$ and the 
area cut of $\ge150$\,arcsec$^2$, translates (using the SSP models) into a mass detection limit increasing with age,
shown with a black line in Fig. \ref{fig:ages} for a foreground reddening $\ebv=0.22$\,mag.
The actual mass limit of the sample is higher because most sources have a higher reddening.
As can be expected, the detection limit causes incompleteness for low masses at old ages.

For comparison, we also estimated the masses of the FUV-selected SF regions from 
resolved stellar  photometry.
The mass of each SF region  was derived by extrapolating the number of stars  above
10\,\Msun\ (corresponding to about B2V, and chosen with \MV\,$<-2.5$  and $(B-V)_0 <
-0.2$) and up to the
most massive star still on the MS, to the interval $0.1-100$\,\Msun.
We assumed an IMF with $\alpha=2.3$ in the range $0.5 < M < 80 $\,\Msun, and
$\alpha=1.3$ for $0.1 < M < 0.5 $\,\Msun\ after~\citet{Kroupa01}.
The masses derived from the H-R diagrams agree with those from 
integrated UV photometry  for SF regions younger than 10\,Myr.
 For older regions, the H-R diagrams give lower masses than the
 integrated measurements, by up to a factor of 20, probably because the most massive
stars have evolved. 
 The most discrepant regions 
have large areas and are in the main body of the galaxy, where the background is higher;
 they may be the result
of merging of nearby regions expanding with age.  

The possible contribution  by foreground MW red
dwarfs to  the integrated  $GALEX$ flux of the SF regions 
was also estimated, since the density of 
foreground stars is significant in the direction of NGC\,6822.
We used our stellar model grids to estimate the possible contribution to the
FUV and NUV fluxes from foreground  stars
 of intermediate colors ($0.1 < (B-V)_0 < 1.2$
and $\MV<-2$, see for example Figure 6 of Bianchi \& Efremova 2006). 
 The derived potential effect on the FUV-NUV color
is very small ($-0.014$ mag on average)
and does not affect the results.

The main concerns to be addressed when deriving ages and masses of SF complexes
by comparison with SSP model colors are: 1) the degree of coevality of the stellar
complex and  applicability of the SSP assumption; 2) stochastic effects from the 
top-IMF small number statistics. The latter affects both the analysis of integrated measurements
 and of resolved stellar counts.  However, it affects only the small mass clusters.
As we explained above, we restricted our analysis sample so to  include sources with
$\geq$5 massive stars, in order to minimize problems of stochasticity.
More importantly, we do not interpret our results in terms of mass
distributions of individual clusters; 
instead, we add the masses of SF regions in broad age bins (next section) in order to obtain the
total stellar mass formed at different epochs.
In this way, we derive  information on global star formation with broad time-resolution, and stochastic effects
on individual cluster masses average out. 
As for the assumption of ``SSP'' (or instantaneous star-formation of each region), 
we tested the results by using also models with
exponential SFH,  decaying  over short time scales:  the results showed no appreciable difference.  
The measured FUV-NUV color of most sources is incompatible
with CSP (``continuous star-formation stellar populations'') model colors of any age, 
ruling out the  CSP assumption often used to derive global galaxy SFR, as not applicable to the individual
SF regions in our sample.
Strict coevality is not observed even in {\it bona fide} globular clusters, the epitome of ``single age''
stellar population, therefore some degree of uncertainty is carried in all works by this assumption, which
is however the most compatible with the observed properties of our young populations sample.

\section{Results and Discussion}
\label{sec:results}

\subsection{Recent star formation from UV fluxes}
\label{sec:sfr}

 The  SF regions have ages $\lesssim400$\,Myr, as derived in the previous section from UV photometry,
due to the FUV selection,
and their masses  range from $2.0\times10^2$\,\Msun\ to $1.5\times10^6$\,\Msun,
when individual extinction correction is applied to the sources, as explained 
in Sec.~\ref{sec:parameters}.

\begin{figure}[htb]
\hskip -1.cm
\epsscale{1.}
\plotone{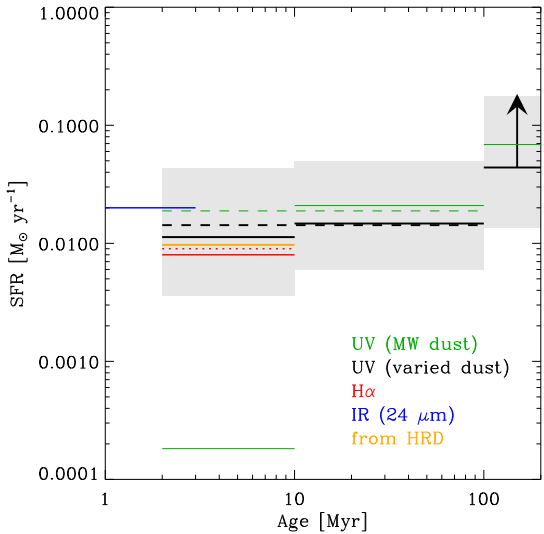}
\vskip -0.5cm
\caption{
The average SFR of NGC6822 over recent time intervals, 
derived by adding the masses of the FUV-defined SF regions 
within each epoch (solid lines; the dashed line shows the average over
the wider range). 
The black lines show results 
obtained with reddening corrections as described in the text; gray boxes
show the uncertainties, taking into account extinction spread within each region and photometric errors. 
Green lines show results obtained assuming 
 MW-type ($\RV=3.1$) extinction for all sources. 
The red lines show the SFR estimated from \Ha\ in this paper (solid line), and 
by \citet{Hunter04} (dotted line).
The blue line indicates the SFR derived from 24\,\micron\ emission, and the 
yellow line  shows the SFR derived from resolved
stellar photometry of the youngest regions.
}
\label{fig:sfr} 
\end{figure}

We added the UV-derived masses of the SF regions  in 
four age ranges, to estimate the average SFR within these time intervals. 
We find: 
SFR\,$=1\times10^{-2}$\,\Msun\,yr$^{-1}$ ($2-10$ Myr), 
SFR\,$=1.5\times10^{-2}$\,\Msun\,yr$^{-1}$ ($10-100$ Myr),
SFR\,$=4.4\times10^{-2}$\,\Msun\,yr$^{-1}$ ($100-200$ Myr), and 
SFR\,$=1.4\times10^{-2}$\,\Msun\,yr$^{-1}$ over the whole range $2-100$\,Myr.
The results are shown in Figure~\ref{fig:sfr}; the uncertainties, shown as gray boxes,
take into account the photometric errors
and the \ebv\ scatter within the SF regions, which is typically one order of magnitude
larger than the photometric errors. 
The uncertainties are up to a factor of four of the derived values.
Additional (smaller) uncertainties may arise from the  assumption that the stellar population in each SF region
has formed in a single burst, and from the adopted IMF.  Coevality is supported by the 
HR diagrams of seven young SF regions, studied with $HST$ (Bianchi et al. 2001,
Bianchi \& Efremova 2006), but it may be more questionable for older larger  regions, 
which may result from merging of subcomponents dissolving with time. 
Stochastic effects for low mass complexes, as
discussed previously, could affect individual masses, but average out
when masses of several clusters are summed in broad age bins. 
Moreover, the total mass is dominated by the most massive SF regions. 

The effect of the extinction curve used to correct the UV color can be appreciated in
Fig.5 where we also show results derived
assuming the total reddening to be from  MW-type dust  
 for all sources (green lines), as was assumed e.g.  by
\citet{Wyder07} and \citet{Hunter10}.
The resulting SFR is significantly lower in the most recent epoch,
because the most massive SF regions are shifted to older ages when we use this extinction curve.

For older ages, an incompleteness
at low masses sets in, driven by our flux detection limit (see Fig.\ref{fig:ages}).
Our FUV-flux  threshold translates (using our SSP models) into mass limits of 
$\sim60$\,\Msun\ , $\sim170$\,\Msun\ , 
and 2900\,\Msun\ at $5$\,Myr,$10$\,Myr and $100$\,Myr respectively,
if only foreground extinction were present. 
The actual limit is higher since most
young stellar populations have additional internal extinction. 
The average detection limit in the $2-10$\,Myr bin is about 100\,\Msun.
The  cluster mass function derived by \citet{Lada03} for embedded
clusters in the solar vicinity has an exponent $\approx -2$ down to
cluster masses of about $50$\,\Msun, then it drops.
If Lada \& Lada's results for solar vicinity were applicable to this dwarf irregular galaxy, 
 the contribution of small mass clusters ($50 - 100$\,\Msun)
to the total mass would be about 10\% in this age bin, where we detect clusters with mass up to 
$4\times10^{4}$\,\Msun. However, star formation in NGC\,6822 is patchy and may not resemble that of a massive
spiral galaxy (Bianchi et al. 2001, Bianchi et al.2010, 2011); in any case 
this estimate should be regarded as an upper limit since
the majority  ($\sim90$\%) of the embedded clusters  are expected 
to merge and become part of larger OB associations or field stars
before they are 10\,Myr\ old  \citep{Lada03}.
At older ages, the  mass function for embedded clusters is no longer applicable:
according to \citet{Lada03},  after 100 Myr 94\% of the embedded individual clusters have merged into 
large OB associations (if there was no cluster disruption, and if the above mass function
were applicable, our detection limit would miss 45\% of the clusters' mass at this epoch:
this approximate figure can be taken as a conservative upper limit). 
The sum of the masses of our SF regions with ages between
 $100-200$ Myr  yields SFR\,${\geq}0.044$\,\Msun\,yr$^{-1}$ in this period,  in
agreement with the study of stellar populations by \citet{Gallart96}, who
found SFR\,$=0.04$\,\Msun\,yr$^{-1}$ in this epoch (adopting $\ebv =0.24$\,mag). 
However,  our value should be considered as a lower limit on SFR because 
the FUV detection threshold corresponds to a high mass detection-limit for old populations.
In addition, the uncertainty factors in the UV-based method, discussed previously, become
significant at ages older than $\sim$100 Myr. 

 We point out that our photometry of  individual 
regions aims at isolating young SF complexes, and
 the older, diffuse galaxy population is subtracted from the flux.
Therefore, it would not be appropriate to compare the total flux from the
SF regions photometry with integrated galaxy models assuming a global SFH (see also Section 2.3).

\subsection{The Very Recent Star Formation from \Ha\ and far-IR Measurements}
\label{sec:sfrha}

We also estimated the  SFR of NGC6822 in very recent epochs 
from the \Ha\ emission (Section \ref{sec:haphot}). 
The total  \Ha\ flux from HII regions measured in this paper (the sum of the flux of the
sources in Table~\ref{tbl:int_phha})
 is $F_{\Ha}=1.8\pm 0.2\times 10^{-11}$ erg \persec\,\percmsq\ . 
After correcting the flux of  the individual sources for reddening, using \ebv\ 
values from the associated or nearest FUV sources, 
and A$_{\Ha}/\ebv=2.5$, the total unreddened flux is 
 $F_{\Ha}(unreddened)= 4\pm 1 \times 10^{-11}$ erg \persec\,\percmsq ,
corresponding to a luminosity of  $\log L_{\Ha}=39.1\pm 0.2$ 
 erg\,\persec. The uncertainty takes into account photometric errors and 
\ebv\ scatter within individual SF regions.
Most of the  \Ha\ emission ($\sim70$\%) comes from \Ha\ regions \# 5, 15, 26 
(see Fig.~\ref{fig:fuv_ha} and Table~\ref{tbl:int_phha}).
These include the H\ II regions Hubble~V and Hubble~X, where an excellent agreement was 
found (under the assumption of optically thick gas) between the \Ha\ luminosity and 
the number of ionizing photons estimated using resolved $HST$ photometry (Bianchi et al. 2001,
Bianchi \& Efremova 2006) and $VLT$ spectroscopy (B. Efremova et al. 2011, in preparation).
The  \Ha\ luminosity translates into SFR\,$=0.008\pm0.003$\,\Msun\,yr$^{-1}$
using the calibration by Panuzzo et al. (2003), which is based on the same SSP models we used
to analyze the UV fluxes, the difference between the case (models) with and without dust 
being less than the uncertainty. Using other calibrations we obtain similar results:
for example SFR\,$=0.01\pm0.003$\,\Msun\,yr$^{-1}$ if we
 use the calibration by 
\citet{Hirashita03}, with $f$=1. 
Among our \Ha-defined HII regions there is one (FUV source EB-FUV \#52, \Ha\ source EB-\Ha\ \#25) 
which includes only two blue stars 
(an O-type star and an early B-type supergiant, according to our $VLT$ spectroscopy).
This gives an indication of the sensitivity of our \Ha\ source detection.  
In general, \Ha-based SFRs should always be regarded as a lower limit, due to the possibility of
leakage of ionizing photons. 

 \Ha\ emission traces  the hottest, most  massive stars, and this estimate is in good agreement with
the UV-derived SFR in the recent $2-10$\,Myr, as we would expect.
Our result is also in agreement with 
previous estimates of SFR based on \Ha\ luminosity: SFR$=0.01$\,\Msun\,yr$^{-1}$ by \citet{Hunter04}, 
and SFR\,$=0.016$\,\Msun\,yr$^{-1}$ by \citet{Cannon06}.

Star formation more recent than $1-2$\,Myr is embedded in dust and
detectable by 24$\micron$ dust emission rather than in the UV, where the stellar flux is still heavily obscured.
We estimated  
the SFR from the 24\,\micron\ emission using the $Spitzer$/MIPS measurements by \citet{Cannon06}, and 
the second term in equation D10 of \citet{Leroy08}.
We derived SFR(24\,\micron)\,$= 0.02$\,\Msun\,yr$^{-1}$; this value is shown with a blue line in Fig.~\ref{fig:sfr},
over a very short time interval, since IR dust emission typically traces the youngest populations,
where the massive stars have not yet dissipated the dust of the parent cloud. Therefore, the 
far-IR detected star formation should complement the stellar mass of young populations detected from FUV measurements. 

\begin{figure*}[htb]
\hskip -1.cm
\epsscale{1.8}
\plotone{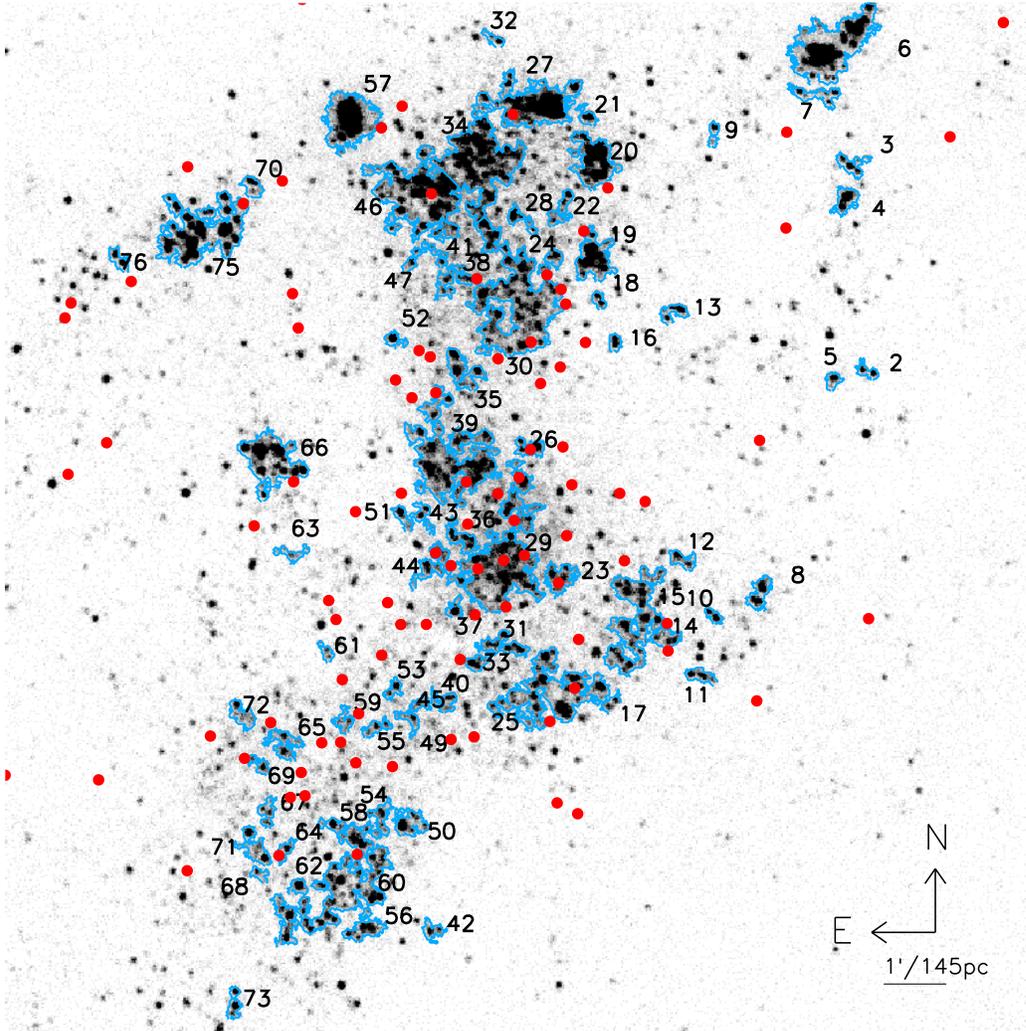}
\vskip -0.5cm
\caption{
The distribution of FUV-defined SF regions (blue contours)  
 compared with the location of known Cepheids in NGC\,6822
from the catalog of \citet{Pietrzynski04} (red dots),
shown over the $GALEX$ FUV image.
}
\label{fig:cepheids} 
\end{figure*}

\section{Summary and Conclusions}
\label{sec:summary}

 We have defined regions of recent star formation in NGC6822 from  $GALEX$ wide-field
FUV imaging, and derived ages and masses from their FUV, NUV photometry compared with
SSP model populations. 
UV light is a good tracer of stellar populations up to a few hundred million years old.
Extinction by interstellar dust has been estimated in each SF region from 
resolved photometry of the stars it contains: with an  average value of \ebv=0.37mag,
it exceeds in most cases the foreground reddening of \ebv=0.22mag.
The characteristics of the internal (within NGC6822) selective extinction at UV wavelengths 
have been explored by comparing
results from integrated UV photometry with H-R diagrams from high-resolution stellar  photometry
available for a number of well studied SF  regions. We found that a UV-steep, non MW-type extinction is preferable for 
high surface brightness SF regions, and adopt it for sources with surface
brightness $<$25~mag~arcsec$^{-2}$. MW-type extinction with $\Rv=3.1$ seems adequate for  
less compact regions, which generally tend to be older. 
Such {\it ad hoc} criterion, suggested by our study for this particular case, 
will be verified over a larger sample of Local Group galaxies
with new multi-band $HST$ imaging (e.g. Bianchi et al. 2010, 2011). 
This study has shown quantitatively that large variations of dust characteristics
as a function of environment, and related to star-formation intensity, exist within a single galaxy. The significant effect 
that the extinction correction bears on SFR derived from UV fluxes (Fig. \ref{fig:sfr}) highlights the limitations of 
integrated SFR recipes, if internal extinction is not properly accounted for. 

We avoided commonly used methods 
for deriving \ebv\ from the ratio of FUV
to 24\micron\ fluxes (e.g. Burgarella et al. 2005), because most far-IR emitting regions in this galaxy are
clearly not co-located with the FUV emitting regions (e.g. Bianchi 2007). 
UV and far-IR bands trace  different populations, and 
using the flux  ratio would consequently overcorrect the FUV fluxes.  
As can be seen in the 24\,\micron\ $Spitzer$ image of NGC\,6822 published by \citet{Cannon06},
most peaks of IR emission trace  \Ha\ emission sites, with the exception of their region 11,
which is not a source of enhanced \Ha\ emission. 
The 24\,\micron\ emission 
originates from dust heated by newly formed stars, still embedded 
in their parental clouds.
The FUV-bright regions are more uniformly spread, because populations older than $\sim10$\,Myr, 
no longer associated with dust nor significantly ionizing gas, still emit
detectable FUV flux; only the youngest, most compact regions are bright in both FUV and \Ha .

We estimated the total stellar mass formed in recent time intervals, by summing the masses
of individual SF regions of corresponding ages. We derive an average SFR=$1.4\times10^{-2}$\,\Msun\,yr$^{-1}$ 
over the past 100 Myr. For older ages the FUV-detection becomes incomplete and our method less robust,
due to dissolving and possible merging of aging SF complexes.  The uncertainty on SFR due to photometric errors,
and extinction correction (the major factor) is very large (shown in Fig. 5). 
The overall level of star-formation activity may not be significantly variable in the last 100Myr, if we
add to the FUV-detected  young populations the embedded star-formation component, 
traced by far-IR  emission from the dust  which  extinguishes the FUV flux.
We found, similarly to Kang et al (2009), that the \Ha\ SFR estimate is a 
good measurement of the recent ($<10$\,Myr) star-formation as assessed by the FUV imaging, when it is 
concentrated in bright compact sources, which are likely to be optically thick. 

Finally, we examine the location of the known Cepheids in NGC\,6822 using the catalog by \citet{Pietrzynski04}. 
Such stars are the evolved descendants of populations formed mostly in earlier epochs than those sampled by this work. 
A few are within our FUV-source contours but mostly they avoid the 
FUV-bright regions and 
follow instead the optical appearance of the galaxy (Figure~\ref{fig:cepheids}). 
While the youngest, FUV-bright SF regions are mostly found in the northern third of the galaxy,
the Cepheids populate more uniformly the middle and southern part of the galaxy.  
We estimated their ages from the period-age relation derived by \citet{Efremov03} for Cepheids
of similar metallicity (in the LMC). According to this relation, 85\% of the Cepheids
in NGC\,6822 are older than 70\,Myr.
However, the period-age relation is constrained by
very few data points at young ages (see Fig.3 of Efremov 2003), and we consider the 
overall spatial distribution more informative than individual ages.

\paragraph{Acknowledgments}
We thank Philip Massey for very helpful clarifications on the calibration 
of the \Ha\ image, 
 Alin Tolea for initial discussions about the source contour definition, and
the anonymous referee for valuable comments.

The $GALEX$ data presented in this paper were obtained from the 
Multimission Archive at the Space Telescope Science Institute (MAST). 
STScI is operated by the Association of Universities for Research in 
Astronomy, Inc., under NASA contract NAS5-26555. Support for MAST for 
non-HST data is provided by the NASA Office of Space Science via grant 
NAG5-7584 and by other grants and contracts.
$GALEX$ ({\it The Galaxy Evolution Explorer}) is a NASA Small Explorer, launched in 
April 2003.
The \Ha\ image used in this paper was obtained by \citet{Massey07b}
as a part of the Survey of Local Group Galaxies Currently Forming Stars
and downloaded from http://www.archive.noao.edu/nsa/.

We gratefully acknowledge NASA's support for construction, operation,
and science analysis of the $GALEX$ mission,
developed in cooperation with the Centre National d'Etudes Spatiales
of France and the Korean Ministry of Science and Technology.
S.-C. R. is supported by the NRF of Korea to the Center for Galaxy Evolution Research.

\renewcommand{\arraystretch}{1.}
\vskip -10cm
\begin{deluxetable}{r@{\ \ \ }|ccr@{\ /}rcccccl}
\tabletypesize{\tiny}
\tablewidth{0pt}
\textheight  25cm
\tablecaption{FUV-selected SF regions
\label{tbl:int_ph} }
\tablehead{
 \multicolumn{1}{c}{\#\tablenotemark{a}}  & \multicolumn{1}{c}{$\alpha$ (J2000)\tablenotemark{b}}  & \multicolumn{1}{c}{$\delta$ (J2000)\tablenotemark{b}} & \multicolumn{2}{c}{Area} & \multicolumn{1}{c}{FUV\tablenotemark{c}} & \multicolumn{1}{c}{NUV\tablenotemark{c}}  & \multicolumn{2}{c}{Background fraction\tablenotemark{d}}  & \multicolumn{1}{c}{$\ebv$\tablenotemark{e}} & \multicolumn{1}{c}{Comment\tablenotemark{f}} \\
\multicolumn{1}{c}{}  & \multicolumn{1}{c}{}  & \multicolumn{1}{c}{} & \multicolumn{2}{c}{[arcsec$^2$/pc$^2$]} & \multicolumn{1}{c}{[AB mag]} & \multicolumn{1}{c}{[AB mag]}  & \multicolumn{1}{c}{FUV}  & \multicolumn{1}{c}{NUV}  & \multicolumn{1}{c}{[mag] and $1\sigma$ scatter} & \multicolumn{1}{c}{} } 
\startdata
EB-FUV   1\   &   19 44 14.46 &   -14 46 39.1 &       190 &      1110 & $ 18.81\pm 0.02$ & $ 18.59\pm 0.01$ & 0.08 & 0.13 & $ 0.50\pm0.12$ &  \\
EB-FUV   2\tablenotemark{g}  &   19 44 31.00 &   -14 47 23.5 &       190 &      1090 & $ 19.70\pm 0.03$ & $ 19.73\pm 0.02$ & 0.21 & 0.30 & $ 0.35\pm0.07$ &  \\
EB-FUV   3\tablenotemark{g}  &   19 44 31.63 &   -14 44 04.0 &       340 &      1950 & $ 18.94\pm 0.02$ & $ 19.09\pm 0.01$ & 0.19 & 0.31 & $ 0.35\pm0.17$ & part of OB2 \\
EB-FUV   4\tablenotemark{g}  &   19 44 32.45 &   -14 44 45.2 &       410 &      2350 & $ 18.23\pm 0.01$ & $ 18.35\pm 0.01$ & 0.15 & 0.25 & $ 0.36\pm0.07$ & part of OB2 \\
EB-FUV   5\tablenotemark{g}  &   19 44 33.02 &   -14 47 32.5 &       190 &      1090 & $ 19.66\pm 0.03$ & $ 19.70\pm 0.02$ & 0.20 & 0.31 & $ 0.41\pm0.08$ &  \\
EB-FUV   6\tablenotemark{g}  &   19 44 33.13 &   -14 42 01.7 &      3440 &     19810 & $ 15.67\pm 0.01$ & $ 15.57\pm 0.01$ & 0.12 & 0.16 & $ 0.31\pm0.13$ & OB1 and OB3 \\
EB-FUV   7\   &   19 44 33.95 &   -14 42 57.2 &       450 &      2600 & $ 19.03\pm 0.02$ & $ 19.03\pm 0.02$ & 0.28 & 0.38 & $ 0.31\pm0.20$ &  \\
EB-FUV   8\tablenotemark{g}  &   19 44 37.82 &   -14 50 57.3 &       440 &      2540 & $ 18.66\pm 0.02$ & $ 18.78\pm 0.01$ & 0.24 & 0.36 & $ 0.32\pm0.11$ & OB4 \\
EB-FUV   9\   &   19 44 40.88 &   -14 43 40.0 &       160 &       910 & $ 19.92\pm 0.04$ & $ 20.16\pm 0.05$ & 0.21 & 0.39 & $ 0.44\pm0.15$ &  \\
EB-FUV  10\tablenotemark{g}  &   19 44 40.98 &   -14 51 19.8 &       160 &       900 & $ 19.56\pm 0.03$ & $ 19.57\pm 0.02$ & 0.19 & 0.28 & $ 0.66\pm0.21$ &  \\
EB-FUV  11\   &   19 44 41.70 &   -14 52 17.5 &       300 &      1720 & $ 19.11\pm 0.02$ & $ 19.17\pm 0.01$ & 0.20 & 0.31 & $ 0.36\pm0.11$ &  \\
EB-FUV  12\   &   19 44 42.74 &   -14 50 25.0 &       280 &      1630 & $ 19.22\pm 0.03$ & $ 18.97\pm 0.01$ & 0.23 & 0.30 & $ 0.42\pm0.04$ &  \\
EB-FUV  13\tablenotemark{g}  &   19 44 43.67 &   -14 46 30.3 &       280 &      1630 & $ 18.81\pm 0.02$ & $ 18.72\pm 0.01$ & 0.15 & 0.24 & $ 0.38\pm0.12$ &  \\
EB-FUV  14\   &   19 44 43.82 &   -14 51 36.3 &       360 &      2090 & $ 19.26\pm 0.03$ & $ 19.31\pm 0.02$ & 0.32 & 0.45 & $ 0.40\pm0.18$ &  \\
EB-FUV  15\   &   19 44 46.05 &   -14 51 28.1 &      2700 &     15550 & $ 16.89\pm 0.01$ & $ 16.78\pm 0.01$ & 0.30 & 0.41 & $ 0.35\pm0.14$ &  \\
EB-FUV  16\tablenotemark{g}  &   19 44 47.24 &   -14 46 59.6 &       160 &       920 & $ 19.59\pm 0.03$ & $ 19.46\pm 0.02$ & 0.18 & 0.28 & $ 0.48\pm0.17$ &  \\
EB-FUV  17\   &   19 44 47.91 &   -14 52 40.8 &       190 &      1100 & $ 19.89\pm 0.04$ & $ 19.84\pm 0.03$ & 0.28 & 0.38 & $ 0.40\pm0.05$ &  \\
EB-FUV  18\tablenotemark{g}  &   19 44 48.43 &   -14 46 16.8 &       150 &       870 & $ 19.63\pm 0.03$ & $ 19.64\pm 0.02$ & 0.19 & 0.32 & $ 0.29\pm0.00$ &  \\
EB-FUV  19\tablenotemark{g}  &   19 44 48.90 &   -14 45 30.3 &      1160 &      6680 & $ 17.03\pm 0.01$ & $ 16.93\pm 0.01$ & 0.17 & 0.25 & $ 0.41\pm0.13$ & OB7 \\
EB-FUV  20\tablenotemark{g}  &   19 44 49.36 &   -14 44 00.3 &      1500 &      8620 & $ 16.54\pm 0.01$ & $ 16.59\pm 0.01$ & 0.17 & 0.25 & $ 0.34\pm0.14$ & OB6 \\
EB-FUV  21\tablenotemark{g}  &   19 44 49.52 &   -14 43 15.3 &       200 &      1130 & $ 19.43\pm 0.03$ & $ 19.52\pm 0.02$ & 0.28 & 0.38 & $ 0.30\pm0.15$ &  \\
EB-FUV  22\   &   19 44 50.96 &   -14 44 52.8 &       360 &      2060 & $ 19.07\pm 0.03$ & $ 19.18\pm 0.02$ & 0.35 & 0.49 & $ 0.32\pm0.14$ &  \\
EB-FUV  23\   &   19 44 51.06 &   -14 50 43.1 &       590 &      3380 & $ 18.72\pm 0.03$ & $ 18.73\pm 0.02$ & 0.38 & 0.52 & $ 0.45\pm0.15$ &  \\
EB-FUV  24\   &   19 44 51.53 &   -14 45 39.3 &       200 &      1160 & $ 19.74\pm 0.04$ & $ 19.82\pm 0.03$ & 0.37 & 0.51 & $ 0.31\pm0.12$ &  \\
EB-FUV  25\tablenotemark{g}  &   19 44 52.35 &   -14 52 38.6 &      3480 &     20030 & $ 16.49\pm 0.01$ & $ 16.42\pm 0.01$ & 0.28 & 0.38 & $ 0.40\pm0.16$ & part of OB5 \\
EB-FUV  26\   &   19 44 53.13 &   -14 48 37.1 &       330 &      1930 & $ 18.96\pm 0.02$ & $ 18.97\pm 0.02$ & 0.28 & 0.44 & $ 0.44\pm0.11$ &  \\
EB-FUV  27\tablenotemark{g}  &   19 44 53.39 &   -14 43 02.6 &      2760 &     15900 & $ 15.67\pm 0.01$ & $ 15.55\pm 0.01$ & 0.14 & 0.18 & $ 0.41\pm0.15$ & OB8 \\
EB-FUV  28\   &   19 44 53.44 &   -14 44 59.6 &       270 &      1560 & $ 19.08\pm 0.03$ & $ 19.17\pm 0.02$ & 0.33 & 0.45 & $ 0.32\pm0.12$ &  \\
EB-FUV  29\   &   19 44 54.48 &   -14 50 16.1 &      5060 &     29160 & $ 16.24\pm 0.01$ & $ 16.12\pm 0.01$ & 0.39 & 0.51 & $ 0.42\pm0.18$ &  \\
EB-FUV  30\   &   19 44 54.79 &   -14 46 16.8 &      5400 &     31100 & $ 16.15\pm 0.01$ & $ 16.00\pm 0.01$ & 0.37 & 0.48 & $ 0.38\pm0.18$ &  \\
EB-FUV  31\   &   19 44 55.00 &   -14 51 49.1 &       520 &      3010 & $ 18.54\pm 0.02$ & $ 18.68\pm 0.02$ & 0.30 & 0.46 & $ 0.44\pm0.15$ &  \\
EB-FUV  32\   &   19 44 55.77 &   -14 42 04.8 &       160 &       920 & $ 19.86\pm 0.03$ & $ 20.05\pm 0.02$ & 0.23 & 0.35 & $ 0.22\pm0.02$ &  \\
EB-FUV  33\   &   19 44 56.29 &   -14 52 01.1 &       250 &      1460 & $ 19.28\pm 0.03$ & $ 19.13\pm 0.02$ & 0.31 & 0.38 & $ 0.31\pm0.06$ &  \\
EB-FUV  34\   &   19 44 56.34 &   -14 44 05.6 &      3920 &     22560 & $ 16.07\pm 0.01$ & $ 15.99\pm 0.01$ & 0.31 & 0.39 & $ 0.33\pm0.19$ & OB9 \\
EB-FUV  35\tablenotemark{g}  &   19 44 57.48 &   -14 47 28.8 &       730 &      4180 & $ 18.01\pm 0.02$ & $ 17.74\pm 0.01$ & 0.30 & 0.39 & $ 0.41\pm0.18$ &  \\
EB-FUV  36\   &   19 44 57.89 &   -14 50 04.1 &       410 &      2380 & $ 18.87\pm 0.03$ & $ 18.74\pm 0.02$ & 0.38 & 0.49 & $ 0.33\pm0.13$ &  \\
EB-FUV  37\   &   19 44 58.10 &   -14 51 11.6 &       210 &      1190 & $ 19.56\pm 0.03$ & $ 19.45\pm 0.07$ & 0.34 & 0.45 & $ 0.23\pm0.02$ &  \\
EB-FUV  38\   &   19 44 58.26 &   -14 45 57.3 &       530 &      3040 & $ 18.97\pm 0.03$ & $ 19.09\pm 0.03$ & 0.46 & 0.63 & $ 0.41\pm0.22$ &  \\
EB-FUV  39\   &   19 44 58.36 &   -14 48 41.6 &      4220 &     24320 & $ 16.27\pm 0.01$ & $ 16.05\pm 0.01$ & 0.32 & 0.43 & $ 0.42\pm0.20$ &  \\
EB-FUV  40\   &   19 44 58.97 &   -14 52 41.6 &       290 &      1700 & $ 19.74\pm 0.04$ & $ 19.89\pm 0.03$ & 0.43 & 0.56 & $ 0.32\pm0.13$ &  \\
EB-FUV  41\   &   19 44 59.29 &   -14 45 31.8 &       180 &      1040 & $ 20.09\pm 0.05$ & $ 20.06\pm 0.04$ & 0.44 & 0.57 & $ 0.60\pm0.25$ &  \\
EB-FUV  42\   &   19 44 59.60 &   -14 56 22.1 &       240 &      1410 & $ 19.68\pm 0.03$ & $ 19.35\pm 0.02$ & 0.23 & 0.28 & $ 0.38\pm0.09$ &  \\
EB-FUV  43\   &   19 45 00.07 &   -14 49 47.6 &       260 &      1520 & $ 19.72\pm 0.04$ & $ 19.87\pm 0.03$ & 0.39 & 0.59 & $ 0.47\pm0.14$ &  \\
EB-FUV  44\   &   19 45 00.22 &   -14 50 38.6 &       330 &      1900 & $ 19.26\pm 0.03$ & $ 19.32\pm 0.03$ & 0.33 & 0.48 & $ 0.32\pm0.05$ &  \\
EB-FUV  45\   &   19 45 00.58 &   -14 52 46.8 &       170 &       990 & $ 20.55\pm 0.06$ & $ 20.71\pm 0.06$ & 0.52 & 0.64 & $ 0.41\pm0.08$ &  \\
EB-FUV  46\   &   19 45 00.73 &   -14 44 43.8 &      4170 &     24030 & $ 16.09\pm 0.01$ & $ 16.06\pm 0.01$ & 0.29 & 0.40 & $ 0.30\pm0.19$ & OB11 \\
EB-FUV  47\   &   19 45 00.73 &   -14 45 34.8 &       190 &      1120 & $ 19.91\pm 0.04$ & $ 19.98\pm 0.04$ & 0.38 & 0.55 & $ 0.41\pm0.14$ &  \\
EB-FUV  48\   &   19 45 00.94 &   -14 58 53.6 &       200 &      1140 & $ 19.40\pm 0.02$ & $ 19.43\pm 0.01$ & 0.13 & 0.22 & $ 0.26\pm0.04$ &  \\
EB-FUV  49\   &   19 45 01.05 &   -14 52 58.8 &       230 &      1320 & $ 20.00\pm 0.05$ & $ 19.97\pm 0.03$ & 0.45 & 0.56 & $ 0.40\pm0.10$ &  \\
EB-FUV  50\   &   19 45 01.20 &   -14 54 33.3 &       570 &      3270 & $ 18.42\pm 0.02$ & $ 18.02\pm 0.01$ & 0.26 & 0.28 & $ 0.38\pm0.12$ &  \\
EB-FUV  51\   &   19 45 01.41 &   -14 49 46.8 &       220 &      1250 & $ 19.42\pm 0.03$ & $ 19.32\pm 0.02$ & 0.25 & 0.37 & $ 0.40\pm0.23$ &  \\
EB-FUV  52\tablenotemark{g}  &   19 45 01.93 &   -14 46 55.8 &       170 &       990 & $ 18.98\pm 0.02$ & $ 18.73\pm 0.01$ & 0.16 & 0.22 & $ 0.40\pm0.09$ &  \\
EB-FUV  53\   &   19 45 02.23 &   -14 52 33.3 &       170 &      1010 & $ 19.97\pm 0.04$ & $ 19.63\pm 0.02$ & 0.35 & 0.38 & $ 0.38\pm0.07$ &  \\
EB-FUV  54\   &   19 45 03.02 &   -14 54 33.3 &       400 &      2280 & $ 19.34\pm 0.03$ & $ 19.25\pm 0.02$ & 0.39 & 0.49 & $ 0.40\pm0.14$ &  \\
EB-FUV  55\   &   19 45 03.22 &   -14 53 07.1 &       250 &      1450 & $ 19.75\pm 0.04$ & $ 19.65\pm 0.03$ & 0.39 & 0.48 & $ 0.33\pm0.13$ &  \\
EB-FUV  56\   &   19 45 04.05 &   -14 56 14.6 &       580 &      3360 & $ 18.48\pm 0.02$ & $ 18.33\pm 0.01$ & 0.22 & 0.31 & $ 0.33\pm0.09$ &  \\
EB-FUV  57\tablenotemark{g}  &   19 45 04.77 &   -14 43 25.8 &      2150 &     12400 & $ 15.46\pm 0.01$ & $ 15.48\pm 0.01$ & 0.08 & 0.13 & $ 0.36\pm0.16$ & OB13 \\
EB-FUV  58\   &   19 45 05.03 &   -14 54 45.3 &       770 &      4460 & $ 17.89\pm 0.01$ & $ 17.76\pm 0.01$ & 0.25 & 0.31 & $ 0.32\pm0.11$ &  \\
EB-FUV  59\   &   19 45 05.44 &   -14 52 59.6 &       270 &      1560 & $ 20.06\pm 0.05$ & $ 20.00\pm 0.04$ & 0.48 & 0.57 & $ 0.28\pm0.23$ &  \\
EB-FUV  60\   &   19 45 06.53 &   -14 55 52.8 &      3630 &     20880 & $ 16.56\pm 0.01$ & $ 16.53\pm 0.01$ & 0.29 & 0.38 & $ 0.31\pm0.12$ & OB12 \\
EB-FUV  61\   &   19 45 06.74 &   -14 51 55.1 &       150 &       870 & $ 20.55\pm 0.05$ & $ 20.36\pm 0.24$ & 0.36 & 0.46 & $ 0.31\pm0.15$ &  \\
EB-FUV  62\   &   19 45 08.66 &   -14 55 41.6 &       190 &      1110 & $ 19.09\pm 0.03$ & $ 19.23\pm 0.02$ & 0.20 & 0.31 & $ 0.27\pm0.02$ &  \\
EB-FUV  63\   &   19 45 08.91 &   -14 50 17.6 &       230 &      1310 & $ 20.05\pm 0.04$ & $ 19.97\pm 0.03$ & 0.31 & 0.44 & $ 0.29\pm0.07$ &  \\
EB-FUV  64\   &   19 45 09.32 &   -14 55 04.1 &       210 &      1210 & $ 19.85\pm 0.04$ & $ 20.05\pm 0.03$ & 0.33 & 0.48 & $ 0.40\pm0.27$ &  \\
EB-FUV  65\   &   19 45 09.48 &   -14 53 23.5 &       520 &      3010 & $ 19.06\pm 0.03$ & $ 19.14\pm 0.02$ & 0.39 & 0.51 & $ 0.33\pm0.09$ &  \\
EB-FUV  66\tablenotemark{g}  &   19 45 10.31 &   -14 48 55.0 &      2100 &     12090 & $ 16.49\pm 0.01$ & $ 16.34\pm 0.01$ & 0.17 & 0.23 & $ 0.26\pm0.16$ & OB14 \\
EB-FUV  67\   &   19 45 10.31 &   -14 54 25.8 &       290 &      1670 & $ 19.65\pm 0.04$ & $ 19.82\pm 0.03$ & 0.36 & 0.50 & $ 0.25\pm0.07$ &  \\
EB-FUV  68\   &   19 45 10.94 &   -14 55 26.5 &       150 &       870 & $ 20.51\pm 0.06$ & $ 20.89\pm 0.06$ & 0.40 & 0.59 & $ 0.52\pm0.22$ &  \\
EB-FUV  69\   &   19 45 11.14 &   -14 53 40.0 &       200 &      1130 & $ 19.94\pm 0.04$ & $ 20.02\pm 0.03$ & 0.33 & 0.46 & $ 0.41\pm0.15$ &  \\
EB-FUV  70\   &   19 45 11.33 &   -14 44 27.3 &       270 &      1570 & $ 19.33\pm 0.03$ & $ 19.44\pm 0.08$ & 0.26 & 0.42 & $ 0.22\pm0.10$ &  \\
EB-FUV  71\tablenotemark{g}  &   19 45 11.60 &   -14 54 58.8 &       530 &      3070 & $ 18.52\pm 0.02$ & $ 18.63\pm 0.01$ & 0.27 & 0.39 & $ 0.35\pm0.14$ &  \\
EB-FUV  72\   &   19 45 12.17 &   -14 52 52.0 &       410 &      2370 & $ 19.30\pm 0.03$ & $ 19.37\pm 0.02$ & 0.36 & 0.48 & $ 0.35\pm0.08$ &  \\
EB-FUV  73\   &   19 45 12.80 &   -14 57 34.0 &       260 &      1510 & $ 19.06\pm 0.02$ & $ 19.06\pm 0.01$ & 0.16 & 0.24 & $ 0.26\pm0.04$ &  \\
EB-FUV  74\   &   19 45 13.47 &   -14 58 49.0 &       240 &      1370 & $ 19.52\pm 0.03$ & $ 19.42\pm 0.02$ & 0.21 & 0.28 & $ 0.29\pm0.07$ &  \\
EB-FUV  75\tablenotemark{g}  &   19 45 14.85 &   -14 45 01.8 &      3410 &     19630 & $ 15.72\pm 0.01$ & $ 15.86\pm 0.01$ & 0.14 & 0.24 & $ 0.30\pm0.20$ & OB15 \\
EB-FUV  76\   &   19 45 20.28 &   -14 45 37.0 &       240 &      1370 & $ 19.31\pm 0.03$ & $ 19.22\pm 0.01$ & 0.21 & 0.29 & $ 0.28\pm0.21$ &  \\
EB-FUV  77\   &   19 45 30.58 &   -14 50 05.4 &       310 &      1800 & $ 19.46\pm 0.03$ & $ 19.72\pm 0.02$ & 0.24 & 0.40 & $ 0.23\pm0.25$ &  \\
\enddata
\tablenotetext{a}{Numbers correspond to the blue labels in Fig.~\ref{fig:fuv_ha}.}
\tablenotetext{b}{Coordinates of the centroids of the contours (see Sec.~\ref{sec:sourcedet}).}
\tablenotetext{c}{Photometric errors are estimated as explained in the text, without adding uncertainties of $GALEX$ calibration zero points.}
\tablenotetext{d}{Fraction of background flux over total flux in the aperture.}
\tablenotetext{e}{The median of values derived for the massive stars (earlier than B2V) in each SF region (see Sec.~\ref{sec:parameters}).}
\tablenotetext{f}{Names of OB associations after \citet{Hodge77}.}
\tablenotetext{g}{The FUV source is associated with \Ha\ emission.}
\end{deluxetable}
\normalsize

\renewcommand{\arraystretch}{1.}
\begin{deluxetable}{r|@{\ \ }ccr@{\ / }rr@{\,$\pm$\,}l@{\ }p{8cm}}
\tabletypesize{\tiny}
\tablewidth{0pt}
\tablecaption{\Ha\  emission regions
\label{tbl:int_phha} }
\tablehead{
\multicolumn{1}{c}{ID} & \multicolumn{1}{c}{$\alpha$ (J2000)\tablenotemark{a}}  & \multicolumn{1}{c}{$\delta$ (J2000)\tablenotemark{a}} & \multicolumn{2}{c}{Area} & \multicolumn{2}{c}{\Ha\ Flux}  & \multicolumn{1}{c}{Comment\tablenotemark{b}} \\
\multicolumn{1}{c}{ }  & \multicolumn{1}{c}{ } & \multicolumn{1}{c}{ }  & \multicolumn{2}{c}{[arcsec$^2$/pc$^2$]} & \multicolumn{1}{r}{[$10^{-13}$ ergs\ } & \multicolumn{1}{l}{sec$^{-1}$ cm$^{-2}$]}  & \multicolumn{1}{c}{} } 
\startdata
EB-\Ha\  1 &   19 44 30.89 &   -14 47 19.7 &       410 &      2380 &  0.24 &  0.03 &  HK1 \\
EB-\Ha\   2 &   19 44 30.93 &   -14 48 29.8 &       250 &      1420 &  1.77 &  0.11 &  K$\alpha$, KD2e  \\
EB-\Ha\   3 &   19 44 32.29 &   -14 44 14.7 &      4420 &     25440 &  7.92 &  0.19 &  HII, Ho1, Ho3, HK2, HK4D, KD2, KD3  \\
EB-\Ha\   4 &   19 44 32.33 &   -14 47 40.9 &       540 &      3090 &  1.22 &  0.18 &  K$\beta$, KD5e  \\
EB-\Ha\   5 &   19 44 32.90 &   -14 42 02.6 &      5990 &     34510 & 42.01 &  0.36 &  HI, HIII, Ho2,KD1, Ho4,KD4 \\
EB-\Ha\   6 &   19 44 38.72 &   -14 42 32.7 &       380 &      2220 &  0.21 &  0.22 &  HK5D \\
EB-\Ha\   7 &   19 44 38.75 &   -14 51 10.2 &      4850 &     27940 &  6.94 &  1.57 &  Ho5, KD8, KD9, HK6, KD7 \\
EB-\Ha\   8 &   19 44 39.98 &   -14 52 00.3 &       340 &      1950 &  0.17 &  0.03 &  HK7, HK8 \\
EB-\Ha\   9 &   19 44 43.68 &   -14 47 00.2 &       160 &       930 &  0.03 &  0.02 &   \\
EB-\Ha\  10 &   19 44 44.36 &   -14 45 57.9 &      1160 &      6710 &  0.62 &  0.46 &  HK11D, HK12 \\
EB-\Ha\   11 &   19 44 47.46 &   -14 46 57.2 &       810 &      4660 &  0.61 &  0.08 &  HK23 \\
EB-\Ha\  12 &   19 44 47.92 &   -14 51 29.5 &       320 &      1830 &  0.22 &  0.05 &   \\
EB-\Ha\  13 &   19 44 48.44 &   -14 46 12.4 &       540 &      3110 &  0.75 &  0.06 &  Ho7, KD11 \\
EB-\Ha\  14 &   19 44 48.45 &   -14 45 24.9 &       670 &      3840 &  0.43 &  0.38 &  HK22, HK27, HK34 \\
EB-\Ha\  15 &   19 44 49.34 &   -14 43 48.0 &      9500 &     54710 & 50.74 &  1.14 &  HV, Ho6, KD12, KD21, HK13, HK15, HK16, HK17, HK19, HK20, HK21, HK32, HK33, HK35, HK36, HK40, HK42, HK44, HK53, Ho9, Ho11, KD19, KD10 \\
EB-\Ha\  16 &   19 44 49.73 &   -14 52 57.8 &       170 &       970 &  1.05 &  0.05 &  KD13, KD13e \\
EB-\Ha\  17 &   19 44 50.52 &   -14 52 06.9 &      3210 &     18520 &  5.41 &  0.34 &  Ho10, K$\gamma$, KD18 , KD11e \\
EB-\Ha\  18 &   19 44 50.57 &   -14 52 45.9 &       550 &      3150 &  2.81 &  0.27 &  Ho8, KD14, KD15, HK48, KD16, KD17 \\
EB-\Ha\  19 &   19 44 52.54 &   -14 44 11.0 &       300 &      1750 &  0.16 &  0.04 &  HK55D \\
EB-\Ha\  20 &   19 44 55.23 &   -14 42 09.5 &       150 &       880 &  0.08 &  0.02 &   \\
EB-\Ha\  21 &   19 44 57.66 &   -14 47 25.9 &       630 &      3620 &  0.83 &  0.16 &  KD24, HK66, HK67 \\
EB-\Ha\  22 &   19 45 00.34 &   -14 43 40.1 &       300 &      1700 &  0.24 &  0.25 &  HK73 \\
EB-\Ha\  23 &   19 45 01.23 &   -14 54 20.8 &       320 &      1840 &  0.08 &  0.04 &   \\
EB-\Ha\  24 &   19 45 01.98 &   -14 54 00.2 &       190 &      1090 &  0.07 &  0.02 &   \\
EB-\Ha\  25 &   19 45 02.38 &   -14 46 55.9 &       970 &      5570 &  1.32 &  0.08 &   \\
EB-\Ha\  26 &   19 45 03.88 &   -14 43 35.3 &      3980 &     22910 & 37.68 &  3.49 &  HX, HK77, HK78, HK79, HK80, Ho14, KD26 \\
EB-\Ha\  27 &   19 45 05.54 &   -14 57 20.6 &      1270 &      7330 &  1.50 &  0.07 &  KD27, KD28 \\
EB-\Ha\  28 &   19 45 09.73 &   -14 44 39.8 &       640 &      3710 &  0.27 &  0.05 &  HK97D \\
EB-\Ha\  29 &   19 45 10.91 &   -14 45 35.4 &      1210 &      6960 &  0.52 &  0.19 &  HK98D \\
EB-\Ha\  30 &   19 45 10.94 &   -14 48 53.4 &      3130 &     18000 &  3.74 &  0.28 &  Ho15, KD30 \\
EB-\Ha\  31 &   19 45 11.41 &   -14 43 47.7 &       580 &      3340 &  0.35 &  0.18 &   \\
EB-\Ha\  32 &   19 45 11.95 &   -14 54 47.7 &       440 &      2540 &  0.16 &  0.11 &   \\
EB-\Ha\  33 &   19 45 14.07 &   -14 58 50.5 &       160 &       910 &  0.18 &  0.02 &   \\
EB-\Ha\  34 &   19 45 14.27 &   -14 43 47.6 &       260 &      1510 &  0.09 &  0.10 &   \\
EB-\Ha\  35 &   19 45 15.83 &   -14 44 58.4 &      3050 &     17560 &  3.24 &  0.20 &  Ho16, HK105D, HK106, HK107, KD31  \\
EB-\Ha\  36 &   19 45 16.03 &   -14 49 13.5 &       260 &      1470 &  0.13 &  0.03 &   \\
EB-\Ha\  37 &   19 45 17.40 &   -14 44 01.5 &       260 &      1500 &  0.09 &  0.06 &   \\
EB-\Ha\  38 &   19 45 17.89 &   -14 44 24.8 &       180 &      1040 &  0.04 &  0.02 &   \\
EB-\Ha\  39 &   19 45 18.87 &   -14 44 57.0 &       190 &      1100 &  0.09 &  0.19 &  HK108 \\
\enddata
\tablenotetext{a}{Coordinates 
of the "centroids" of the H$\alpha$ contours (see Sec.~\ref{sec:sourcedet}).}
\tablenotetext{b}{Cross-identification with H$\alpha$ sources in previous works:
 HI-X, \citet{Hubble25}; Ho followed by arabic number, \citet{Hodge77}; K followed by a greek letter, \citet{Kinman79}; KD followed by arabic number, \citet{Killen82}; HK followed by arabic number, \citet{Hodge88}}
\end{deluxetable}
\normalsize

\clearpage
\renewcommand{\arraystretch}{1.}
\begin{deluxetable}{r@{\ \ \ }|r@{\extracolsep{3pt}}lr@{\extracolsep{1pt}}@{.}l@{\extracolsep{3pt}}lr@{\extracolsep{3pt}}lr@{\extracolsep{1pt}}@{.}l@{\extracolsep{3pt}}lr@{\extracolsep{3pt}}lr@{\extracolsep{1pt}}@{.}l@{\extracolsep{3pt}}ll}
\tabletypesize{\tiny}
\tablewidth{0pt}
\vskip -10cm
\tablecaption{Ages and Masses of the FUV-defined SF regions assuming different extinction curves
\label{tbl:agmass} }
\tablehead{
 \multicolumn{1}{c}{} &   \multicolumn{5}{c}{MW with Rv=3.1\tablenotemark{b}}  & \multicolumn{5}{c}{LMC\tablenotemark{b}} &  \multicolumn{5}{c}{LMC2\tablenotemark{b}} & \multicolumn{1}{c}{ext. curve}\\
\multicolumn{1}{c}{ID\tablenotemark{a}}  & \multicolumn{2}{c}{Age[Myr]}  & \multicolumn{3}{c}{Mass[$10^3$\Msun]} & \multicolumn{2}{c}{Age[Myr]}  & \multicolumn{3}{c}{Mass[$10^3$\Msun]} & \multicolumn{2}{c}{Age[Myr]}  & \multicolumn{3}{c}{Mass[$10^3$\Msun]} & \multicolumn{1}{c}{adopted\tablenotemark{c}} } 
\startdata
 EB-FUV 1 & $   186$ & $^{+    15} _{-    26}$ & $         240$ & $ $ & $^{+   355.} _{-   143.}$ & $    47$ & $^{+    49} _{-    34}$ & $          46$ & $ $ & $^{+    74.} _{-    28.}$ & $     3$ & $^{+    38} _{-     3}$ & $       1$ & $            9$ & $^{+    3.4} _{-    1.2}$ & LMC\\
 EB-FUV   2 & $    61$ & $^{+     1} _{-     2}$ & $       7$ & $            7$ & $^{+    5.5} _{-    3.2}$ & $    25$ & $^{+    20} _{-    14}$ & $       2$ & $            7$ & $^{+    2.1} _{-    1.2}$ & $     5$  & $^{+    22} _{-     4}$ & $       0$ & $            4$ & $^{+    0.3} _{-    0.2}$ & Rv3.1\\
 EB-FUV  3 & $    28$ & $^{+     1} _{-     7}$ & $       6$ & $            0$ & $^{+   15.9} _{-    3.8}$ & $     5$ & $^{+    21} _{-     5}$ & $       0$ & $            8$ & $^{+    2.2} _{-    0.5}$ & $     2$ & $^{+    25} _{-     2}$ & $       0$ & $            7$ & $^{+    2.1} _{-    0.4}$ & Rv3.1\\
 EB-FUV   4 & $    37$ & $^{+     1} _{-     2}$ & $          18$ & $ $ & $^{+    12.} _{-     7.}$ & $     6$ & $^{+    12} _{-     2}$ & $       1$ & $            7$ & $^{+    1.3} _{-    0.8}$ & $     2$ & $^{+     3} _{-     2}$ & $       1$ & $            3$ & $^{+    1.1} _{-    0.6}$ & LMC\\
 EB-FUV   5 & $    56$ & $^{+     2} _{-     5}$ & $          11$ & $ $ & $^{+    10.} _{-     5.}$ & $    11$ & $^{+    17} _{-     7}$ & $       1$ & $            6$ & $^{+    1.5} _{-    0.8}$ & $     2$ & $^{+     4} _{-     2}$ & $       0$ & $            6$ & $^{+    0.6} _{-    0.3}$ & Rv3.1\\
 EB-FUV   6 & $   123$ & $^{+     1} _{-     8}$ & $         560$ & $ $ & $^{+   949.} _{-   278.}$ & $    76$ & $^{+    45} _{-    41}$ & $         320$ & $ $ & $^{+   596.} _{-   166.}$ & $    46$ & $^{+    75} _{-    43}$ & $         180$ & $ $ & $^{+   361.} _{-    96.}$ & SMC\\
 EB-FUV   7 & $    71$ & $^{+     1} _{-     8}$ & $          13$ & $ $ & $^{+    46.} _{-     6.}$ & $    44$ & $^{+    25} _{-    40}$ & $       7$ & $            5$ & $^{+   30.2} _{-    3.9}$ & $    19$ & $^{+    51} _{-    19}$ & $       2$ & $            6$ & $^{+   11.6} _{-    1.4}$ & Rv3.1\\
 EB-FUV   8 & $    37$ & $^{+     1} _{-     2}$ & $       8$ & $            8$ & $^{+   11.5} _{-    4.7}$ & $    13$ & $^{+    23} _{-     9}$ & $       2$ & $            4$ & $^{+    3.4} _{-    1.3}$ & $     4$ & $^{+    32} _{-     4}$ & $       0$ & $            6$ & $^{+    0.9} _{-    0.4}$ & Rv3.1\\
 EB-FUV   9 & $     8$ & $^{+     1} _{-     3}$ & $       1$ & $            0$ & $^{+    2.1} _{-    0.7}$ & - &  & - & &  & - &  & - & &  & Rv3.1\\
 EB-FUV  10 & $    46$ & $^{+    17} _{-    28}$ & $          66$ & $ $ & $^{+   257.} _{-    53.}$ & - &  & - & &  & - &  & - & &  & Rv3.1\\
 EB-FUV  11 & $    51$ & $^{+     1} _{-     3}$ & $          12$ & $ $ & $^{+    15.} _{-     7.}$ & $    16$ & $^{+    28} _{-    12}$ & $       2$ & $            8$ & $^{+    4.0} _{-    1.6}$ & $     3$ & $^{+    33} _{-     3}$ & $       0$ & $            4$ & $^{+    0.7} _{-    0.3}$ & Rv3.1\\
 EB-FUV  12 & $   223$ & $^{+     6} _{-     8}$ & $         120$ & $ $ & $^{+    42.} _{-    32.}$ & $    90$ & $^{+    25} _{-    20}$ & $          37$ & $ $ & $^{+    14.} _{-    10.}$ & $    26$ & $^{+    21} _{-    16}$ & $       8$ & $            6$ & $^{+    3.5} _{-    2.5}$ & Rv3.1\\
 EB-FUV 13 & $   114$ & $^{+     2} _{-    14}$ & $          48$ & $ $ & $^{+    72.} _{-    29.}$ & $    48$ & $^{+    44} _{-    33}$ & $          18$ & $ $ & $^{+    29.} _{-    11.}$ & $    11$ & $^{+    63} _{-    11}$ & $       3$ & $ 0$ & $^{+    5.3} _{-    1.9}$ & LMC\\
 EB-FUV  14 & $    53$ & $^{+     1} _{-     9}$ & $          14$ & $ $ & $^{+    42.} _{-    11.}$ & $    11$ & $^{+    42} _{-    10}$ & $       2$ & $            2$ & $^{+    7.2} _{-    1.7}$ & $     2$ & $^{+    51} _{-     2}$ & $       0$ & $            7$ & $^{+    2.7} _{-    0.6}$ & Rv3.1\\
 EB-FUV  15 & $   129$ & $^{+     1} _{-    14}$ & $         260$ & $ $ & $^{+   492.} _{-   164.}$ & $    64$ & $^{+    64} _{-    44}$ & $         120$ & $ $ & $^{+   244.} _{-    76.}$ & $    27$ & $^{+   101} _{-    26}$ & $          43$ & $ $ & $^{+    98.} _{-    29.}$ & Rv3.1\\
 EB-FUV  16 & $   129$ & $^{+    14} _{-    34}$ & $          58$ & $ $ & $^{+   154.} _{-    42.}$ & $    30$ & $^{+    58} _{-    27}$ & $          11$ & $ $ & $^{+    33.} _{-     8.}$ & $     2$ & $^{+    51} _{-     2}$ & $       1$ & $ 0$ & $^{+    3.2} _{-    0.7}$ & Rv3.1\\
 EB-FUV  17 & $    90$ & $^{+     2} _{-     3}$ & $          16$ & $ $ & $^{+     7.} _{-     5.}$ & $    33$ & $^{+    13} _{-    15}$ & $       5$ & $ 0$ & $^{+    2.5} _{-    1.7}$ & $     4$ & $^{+    10} _{-     2}$ & $       0$ & $            4$ & $^{+    0.2} _{-    0.2}$ & Rv3.1\\
 EB-FUV  18 & $    67$ & $^{+     1} _{-     1}$ & $       6$ & $ 0$ & $^{+    6.7} _{-    2.5}$ & $    46$ & $^{+    20} _{-    28}$ & $       3$ & $            9$ & $^{+    4.9} _{-    1.7}$ & $    27$ & $^{+    39} _{-    24}$ & $       2$ & $            0$ & $^{+    2.8} _{-    0.9}$ & Rv3.1\\
 EB-FUV  19 & $   118$ & $^{+     5} _{-    18}$ & $         330$ & $ $ & $^{+   546.} _{-   204.}$ & $    44$ & $^{+    45} _{-    35}$ & $         100$ & $ $ & $^{+   192.} _{-    68.}$ & $     5$ & $^{+    58} _{-     5}$ & $       7$ & $            6$ & $^{+   15.4} _{-    5.1}$ & LMC2\\
 EB-FUV  20 & $    54$ & $^{+     1} _{-     5}$ & $         110$ & $ $ & $^{+   217.} _{-    68.}$ & $    22$ & $^{+    31} _{-    18}$ & $          39$ & $ $ & $^{+    81.} _{-    24.}$ & $     5$ & $^{+    48} _{-     5}$ & $       6$ & $            2$ & $^{+   14.3} _{-    4.0}$ & LMC2\\
 EB-FUV  21 & $    45$ & $^{+     1} _{-     2}$ & $       4$ & $            6$ & $^{+    9.9} _{-    2.1}$ & $    22$ & $^{+    22} _{-    18}$ & $       2$ & $ 0$ & $^{+    4.6} _{-    0.9}$ & $     6$ & $^{+    38} _{-     6}$ & $       0$ & $            4$ & $^{+    1.0} _{-    0.2}$ & Rv3.1\\
 EB-FUV  22 & $    40$ & $^{+     1} _{-     4}$ & $       6$ & $            7$ & $^{+   12.6} _{-    3.6}$ & $    15$ & $^{+    24} _{-    12}$ & $       1$ & $            8$ & $^{+    3.9} _{-    1.0}$ & $     4$ & $^{+    35} _{-     4}$ & $       0$ & $            5$ & $^{+    1.1} _{-    0.3}$ & Rv3.1\\
 EB-FUV  23 & $    64$ & $^{+     4} _{-    12}$ & $          43$ & $ $ & $^{+    91.} _{-    29.}$ & $     8$ & $^{+    36} _{-     6}$ & $       3$ & $            7$ & $^{+    8.5} _{-    2.6}$ & - &  & - & &  & Rv3.1\\
 EB-FUV  24 & $    46$ & $^{+     1} _{-     2}$ & $       4$ & $ 0$ & $^{+    5.9} _{-    2.0}$ & $    22$ & $^{+    24} _{-    17}$ & $       1$ & $            6$ & $^{+    2.6} _{-    0.8}$ & $     6$ & $^{+    40} _{-     6}$ & $       0$ & $            3$ & $^{+    0.5} _{-    0.2}$ & Rv3.1\\
 EB-FUV  25 & $   100$ & $^{+     2} _{-    15}$ & $         410$ & $ $ & $^{+   977.} _{-   290.}$ & $    39$ & $^{+    53} _{-    34}$ & $         140$ & $ $ & $^{+   366.} _{-   101.}$ & $     5$ & $^{+    78} _{-     5}$ & $          11$ & $ $ & $^{+    32.} _{-     8.}$ & Rv3.1\\
 EB-FUV  26 & $    64$ & $^{+     3} _{-     8}$ & $          32$ & $ $ & $^{+    42.} _{-    18.}$ & $    11$ & $^{+    27} _{-     8}$ & $       3$ & $            9$ & $^{+    5.5} _{-    2.3}$ & - &  & - & &  & Rv3.1\\
 EB-FUV  27 & $   132$ & $^{+     4} _{-    21}$ & $        1200$ & $ $ & $^{+  2570.} _{-   822.}$ & $    50$ & $^{+    66} _{-    39}$ & $         400$ & $ $ & $^{+   932.} _{-   280.}$ & $     8$ & $^{+    88} _{-     8}$ & $          45$ & $ $ & $^{+   115.} _{-    33.}$ & LMC2\\
 EB-FUV  28 & $    44$ & $^{+     1} _{-     2}$ & $       7$ & $            4$ & $^{+   11.0} _{-    3.9}$ & $    18$ & $^{+    26} _{-    13}$ & $       2$ & $            4$ & $^{+    3.9} _{-    1.3}$ & $     5$ & $^{+    38} _{-     5}$ & $       0$ & $            5$ & $^{+    0.9} _{-    0.3}$ & Rv3.1\\
 EB-FUV  29 & $   130$ & $^{+     6} _{-    30}$ & $         820$ & $ $ & $^{+  2381.} _{-   610.}$ & $    45$ & $^{+    77} _{-    40}$ & $         250$ & $ $ & $^{+   795.} _{-   188.}$ & $     5$ & $^{+   102} _{-     5}$ & $          17$ & $ $ & $^{+    62.} _{-    14.}$ & Rv3.1\\
  EB-FUV 30 & $   154$ & $^{+     1} _{-    26}$ & $         830$ & $ $ & $^{+  2405.} _{-   580.}$ & $    66$ & $^{+    89} _{-    53}$ & $         310$ & $ $ & $^{+  1011.} _{-   224.}$ & $    20$ & $^{+   135} _{-    20}$ & $          77$ & $ $ & $^{+   277.} _{-    57.}$ & Rv3.1\\
  EB-FUV 31 & $    29$ & $^{+     3} _{-    10}$ & $          18$ & $ $ & $^{+    38.} _{-    12.}$ & $     3$ & $^{+    12} _{-     3}$ & $       1$ & $            4$ & $^{+    3.3} _{-    1.0}$ & - &  & - & &  & Rv3.1\\
  EB-FUV 32 & $    17$ & $^{+     1} _{-     1}$ & $       0$ & $            5$ & $^{+    0.1} _{-    0.0}$ & $    17$ & $^{+     0} _{-     3}$ & $       0$ & $            5$ & $^{+    0.1} _{-    0.0}$ & $    17$ & $^{+     0} _{-     6}$ & $       0$ & $            5$ & $^{+    0.1} _{-    0.0}$ & Rv3.1\\
  EB-FUV 33 & $   157$ & $^{+     1} _{-     2}$ & $          28$ & $ $ & $^{+    16.} _{-    10.}$ & $    98$ & $^{+    39} _{-    27}$ & $          16$ & $ $ & $^{+    10.} _{-     6.}$ & $    60$ & $^{+    55} _{-    33}$ & $       9$ & $            0$ & $^{+    6.0} _{-    3.6}$ & Rv3.1\\
  EB-FUV 34 & $   109$ & $^{+     1} _{-    16}$ & $         390$ & $ $ & $^{+  1276.} _{-   223.}$ & $    61$ & $^{+    47} _{-    50}$ & $         200$ & $ $ & $^{+   721.} _{-   118.}$ & $    30$ & $^{+    77} _{-    30}$ & $          90$ & $ $ & $^{+   361.} _{-    55.}$ & Rv3.1\\
 EB-FUV  35 & $   244$ & $^{+     9} _{-    44}$ & $         390$ & $ $ & $^{+  1139.} _{-   291.}$ & $   106$ & $^{+   136} _{-    74}$ & $         130$ & $ $ & $^{+   418.} _{-    99.}$ & $    39$ & $^{+   194} _{-    39}$ & $          40$ & $ $ & $^{+   142.} _{-    31.}$ & Rv3.1\\
 EB-FUV  36 & $   143$ & $^{+     1} _{-    10}$ & $          41$ & $ $ & $^{+    70.} _{-    23.}$ & $    80$ & $^{+    62} _{-    42}$ & $          21$ & $ $ & $^{+    39.} _{-    13.}$ & $    44$ & $^{+    98} _{-    41}$ & $          11$ & $ $ & $^{+    21.} _{-     6.}$ & Rv3.1\\
 EB-FUV  37 & $   129$ & $^{+     1} _{-     1}$ & $       8$ & $            9$ & $^{+    1.5} _{-    0.7}$ & $   122$ & $^{+     6} _{-    12}$ & $       8$ & $            4$ & $^{+    1.5} _{-    0.7}$ & $   114$ & $^{+    14} _{-    23}$ & $       7$ & $            8$ & $^{+    1.5} _{-    0.6}$ & Rv3.1\\
 EB-FUV  38 & $    36$ & $^{+     1} _{-    16}$ & $          13$ & $ $ & $^{+    54.} _{-    10.}$ & $     4$ & $^{+    32} _{-     4}$ & $       1$ & $            0$ & $^{+    5.0} _{-    0.8}$ & - &  & - & &  & Rv3.1\\
 EB-FUV  39 & $   197$ & $^{+     7} _{-    37}$ & $        1500$ & $ $ & $^{+  5370.} _{-  1179.}$ & $    77$ & $^{+   127} _{-    65}$ & $         460$ & $ $ & $^{+  1822.} _{-   367.}$ & $    18$ & $^{+   185} _{-    18}$ & $          83$ & $ $ & $^{+   369.} _{-    68.}$ & Rv3.1\\
 EB-FUV  40 & $    28$ & $^{+     1} _{-     3}$ & $       2$ & $            3$ & $^{+    3.9} _{-    1.2}$ & $     7$ & $^{+    19} _{-     5}$ & $       0$ & $            5$ & $^{+    0.9} _{-    0.3}$ & $     3$ & $^{+    24} _{-     3}$ & $       0$ & $            2$ & $^{+    0.4} _{-    0.1}$ & Rv3.1\\
  EB-FUV 41 & $    64$ & $^{+    20} _{-    31}$ & $          38$ & $ $ & $^{+   215.} _{-    33.}$ & $     3$ & $^{+    40} _{-     3}$ & $       1$ & $            3$ & $^{+    8.6} _{-    1.2}$ & - &  & - & &  & Rv3.1\\
 EB-FUV  42 & $   303$ & $^{+     7} _{-    16}$ & $         100$ & $ $ & $^{+    97.} _{-    49.}$ & $   168$ & $^{+    76} _{-    68}$ & $          40$ & $ $ & $^{+    42.} _{-    21.}$ & $    73$ & $^{+   109} _{-    54}$ & $          15$ & $ $ & $^{+    17.} _{-     8.}$ & Rv3.1\\
 EB-FUV  43 & $    24$ & $^{+     4} _{-     7}$ & $       6$ & $ 0$ & $^{+   11.2} _{-    3.9}$ & $     2$ & $^{+     4} _{-     2}$ & $       0$ & $            8$ & $^{+    1.6} _{-    0.5}$ & - &  & - & &  & Rv3.1\\
 EB-FUV  44 & $    51$ & $^{+     1} _{-     1}$ & $       7$ & $            5$ & $^{+    3.5} _{-    2.4}$ & $    25$ & $^{+    15} _{-    10}$ & $       3$ & $            2$ & $^{+    1.6} _{-    1.1}$ & $     6$ & $^{+    18} _{-     3}$ & $       0$ & $            5$ & $^{+    0.3} _{-    0.2}$ & Rv3.1\\
 EB-FUV  45 & $    24$ & $^{+     2} _{-     4}$ & $       1$ & $            7$ & $^{+    1.4} _{-    0.8}$ & $     3$ & $^{+     3} _{-     2}$ & $       0$ & $            2$ & $^{+    0.2} _{-    0.1}$ & - &  & - & &  & Rv3.1\\
 EB-FUV  46 & $    84$ & $^{+     1} _{-     9}$ & $         220$ & $ $ & $^{+   711.} _{-   100.}$ & $    54$ & $^{+    28} _{-    47}$ & $         130$ & $ $ & $^{+   485.} _{-    64.}$ & $    33$ & $^{+    49} _{-    33}$ & $          76$ & $ $ & $^{+   305.} _{-    38.}$ & Rv3.1\\
 EB-FUV  47 & $    48$ & $^{+     1} _{-     6}$ & $       7$ & $            4$ & $^{+   14.1} _{-    4.9}$ & $     6$ & $^{+    31} _{-     4}$ & $       0$ & $            6$ & $^{+    1.2} _{-    0.4}$ & - &  & - & &  & Rv3.1\\
 EB-FUV  48 & $    60$ & $^{+     1} _{-     1}$ & $       5$ & $            1$ & $^{+    1.8} _{-    1.4}$ & $    48$ & $^{+    11} _{-     8}$ & $       3$ & $            9$ & $^{+    1.5} _{-    1.1}$ & $    39$ & $^{+    20} _{-    22}$ & $       3$ & $            1$ & $^{+    1.3} _{-    0.9}$ & Rv3.1\\
 EB-FUV  49 & $    82$ & $^{+     2} _{-     7}$ & $          12$ & $ $ & $^{+    14.} _{-     7.}$ & $    27$ & $^{+    27} _{-    22}$ & $       3$ & $            5$ & $^{+    4.3} _{-    1.9}$ & $     3$ & $^{+    30} _{-     3}$ & $       0$ & $            3$ & $^{+    0.4} _{-    0.2}$ & Rv3.1\\
 EB-FUV  50 & $   374$ & $^{+     9} _{-    26}$ & $         460$ & $ $ & $^{+   672.} _{-   274.}$ & $   221$ & $^{+   121} _{-    99}$ & $         200$ & $ $ & $^{+   321.} _{-   123.}$ & $   107$ & $^{+   192} _{-    85}$ & $          75$ & $ $ & $^{+   132.} _{-    48.}$ & Rv3.1\\
\tablebreak
 EB-FUV  51 & $   119$ & $^{+     2} _{-    33}$ & $          34$ & $ $ & $^{+   159.} _{-    25.}$ & $    46$ & $^{+    75} _{-    43}$ & $          11$ & $ $ & $^{+    60.} _{-     9.}$ & $     6$ & $^{+   115} _{-     6}$ & $       0$ & $            9$ & $^{+    5.2} _{-    0.7}$ & Rv3.1\\
  EB-FUV 52 & $   226$ & $^{+     8} _{-    17}$ & $         130$ & $ $ & $^{+   129.} _{-    65.}$ & $   100$ & $^{+    64} _{-    42}$ & $          46$ & $ $ & $^{+    49.} _{-    24.}$ & $    39$ & $^{+    63} _{-    34}$ & $          15$ & $ $ & $^{+    17.} _{-     8.}$ & LMC\\
  EB-FUV 53 & $   313$ & $^{+     7} _{-    13}$ & $          80$ & $ $ & $^{+    57.} _{-    33.}$ & $   175$ & $^{+    60} _{-    53}$ & $          32$ & $ $ & $^{+    25.} _{-    14.}$ & $    78$ & $^{+    84} _{-    43}$ & $          12$ & $ $ & $^{+    10.} _{-     5.}$ & Rv3.1\\
 EB-FUV  54 & $   113$ & $^{+     4} _{-    17}$ & $          34$ & $ $ & $^{+    64.} _{-    22.}$ & $    44$ & $^{+    49} _{-    36}$ & $          11$ & $ $ & $^{+    24.} _{-     8.}$ & $     5$ & $^{+    69} _{-     5}$ & $       0$ & $            9$ & $^{+    2.0} _{-    0.6}$ & Rv3.1\\
 EB-FUV  55 & $   123$ & $^{+     1} _{-    10}$ & $          15$ & $ $ & $^{+    25.} _{-     9.}$ & $    67$ & $^{+    54} _{-    39}$ & $       7$ & $            6$ & $^{+   14.2} _{-    4.5}$ & $    36$ & $^{+    85} _{-    34}$ & $       3$ & $            7$ & $^{+    7.6} _{-    2.3}$ & Rv3.1\\
  EB-FUV 56 & $   156$ & $^{+     1} _{-     6}$ & $          68$ & $ $ & $^{+    67.} _{-    33.}$ & $    88$ & $^{+    54} _{-    36}$ & $          35$ & $ $ & $^{+    37.} _{-    18.}$ & $    48$ & $^{+    80} _{-    41}$ & $          17$ & $ $ & $^{+    20.} _{-     9.}$ & Rv3.1\\
 EB-FUV  57 & $    64$ & $^{+     1} _{-     8}$ & $         440$ & $ $ & $^{+  1041.} _{-   288.}$ & $    25$ & $^{+    38} _{-    21}$ & $         140$ & $ $ & $^{+   379.} _{-    98.}$ & $     5$ & $^{+    58} _{-     5}$ & $          19$ & $ $ & $^{+    55.} _{-    13.}$ & LMC2\\
 EB-FUV  58 & $   143$ & $^{+     1} _{-     7}$ & $          95$ & $ $ & $^{+   123.} _{-    50.}$ & $    84$ & $^{+    57} _{-    39}$ & $          52$ & $ $ & $^{+    74.} _{-    29.}$ & $    48$ & $^{+    94} _{-    43}$ & $          27$ & $ $ & $^{+    41.} _{-    15.}$ & Rv3.1\\
 EB-FUV  59 & $    97$ & $^{+     1} _{-    11}$ & $       6$ & $ 0$ & $^{+   28.3} _{-    2.2}$ & $    71$ & $^{+    25} _{-    63}$ & $       4$ & $            1$ & $^{+   22.1} _{-    1.6}$ & $    50$ & $^{+    46} _{-    50}$ & $       2$ & $            7$ & $^{+   16.6} _{-    1.1}$ & Rv3.1\\
 EB-FUV  60 & $    84$ & $^{+     1} _{-     4}$ & $         150$ & $ $ & $^{+   230.} _{-    76.}$ & $    51$ & $^{+    32} _{-    33}$ & $          87$ & $ $ & $^{+   142.} _{-    45.}$ & $    27$ & $^{+    56} _{-    25}$ & $          41$ & $ $ & $^{+    72.} _{-    22.}$ & Rv3.1\\
 EB-FUV  61 & $   184$ & $^{+     1} _{-    12}$ & $          11$ & $ $ & $^{+    24.} _{-     6.}$ & $   124$ & $^{+    59} _{-    73}$ & $       6$ & $            6$ & $^{+   15.4} _{-    3.4}$ & $    74$ & $^{+   108} _{-    69}$ & $       3$ & $            6$ & $^{+    9.4} _{-    2.0}$ & Rv3.1\\
 EB-FUV  62 & $    31$ & $^{+     1} _{-     1}$ & $       3$ & $            3$ & $^{+    0.5} _{-    0.5}$ & $    17$ & $^{+     3} _{-     3}$ & $       1$ & $            5$ & $^{+    0.3} _{-    0.2}$ & $     8$ & $^{+     7} _{-     3}$ & $       0$ & $            6$ & $^{+    0.1} _{-    0.1}$ & LMC\\
 EB-FUV  63 & $   110$ & $^{+     1} _{-     2}$ & $       7$ & $            5$ & $^{+    5.3} _{-    3.1}$ & $    76$ & $^{+    32} _{-    25}$ & $       4$ & $            8$ & $^{+    3.7} _{-    2.1}$ & $    50$ & $^{+    57} _{-    34}$ & $       3$ & $ 0$ & $^{+    2.5} _{-    1.3}$ & Rv3.1\\
 EB-FUV  64 & $    16$ & $^{+     1} _{-    10}$ & $       1$ & $            7$ & $^{+   11.7} _{-    1.3}$ & $     3$ & $^{+    13} _{-     3}$ & $       0$ & $            3$ & $^{+    2.6} _{-    0.3}$ & - &  & - & &  & Rv3.1\\
 EB-FUV  65 & $    47$ & $^{+     1} _{-     2}$ & $       8$ & $            6$ & $^{+    8.5} _{-    4.3}$ & $    18$ & $^{+    24} _{-    12}$ & $       2$ & $            6$ & $^{+    2.8} _{-    1.4}$ & $     5$ & $^{+    31} _{-     5}$ & $       0$ & $            5$ & $^{+    0.6} _{-    0.3}$ & Rv3.1\\
 EB-FUV  66 & $   157$ & $^{+     1} _{-     6}$ & $         250$ & $ $ & $^{+   590.} _{-    65.}$ & $   130$ & $^{+    25} _{-    77}$ & $         200$ & $ $ & $^{+   516.} _{-    55.}$ & $   101$ & $^{+    54} _{-    94}$ & $         150$ & $ $ & $^{+   439.} _{-    44.}$ & LMC2\\
 EB-FUV  67 & $    21$ & $^{+     1} _{-     1}$ & $       1$ & $            0$ & $^{+    0.7} _{-    0.2}$ & $    16$ & $^{+     4} _{-    10}$ & $       0$ & $            7$ & $^{+    0.5} _{-    0.1}$ & $    11$ & $^{+     9} _{-     8}$ & $       0$ & $            4$ & $^{+    0.4} _{-    0.1}$ & Rv3.1\\
 EB-FUV  68 & $     2$ & $^{+     1} _{-     2}$ & $       0$ & $            5$ & $^{+    1.9} _{-    0.4}$ & - &  & - & &  & - &  & - & &  & Rv3.1\\
 EB-FUV  69 & $    45$ & $^{+     1} _{-     7}$ & $       6$ & $            8$ & $^{+   14.5} _{-    4.6}$ & $     6$ & $^{+    31} _{-     4}$ & $       0$ & $            5$ & $^{+    1.2} _{-    0.4}$ & - &  & - & &  & Rv3.1\\
 EB-FUV  70 & $    39$ & $^{+     1} _{-     1}$ & $       2$ & $            3$ & $^{+    2.7} _{-    0.0}$ & $    39$ & $^{+     0} _{-    24}$ & $       2$ & $            3$ & $^{+    2.9} _{-    0.0}$ & $    39$ & $^{+     0} _{-    35}$ & $       2$ & $            3$ & $^{+    3.2} _{-    0.0}$ & Rv3.1\\
 EB-FUV  71 & $    40$ & $^{+     1} _{-     5}$ & $          14$ & $ $ & $^{+    26.} _{-     9.}$ & $     9$ & $^{+    30} _{-     6}$ & $       2$ & $            1$ & $^{+    4.4} _{-    1.4}$ & $     3$ & $^{+    36} _{-     3}$ & $       0$ & $            8$ & $^{+    1.8} _{-    0.5}$ & Rv3.1\\
 EB-FUV  72 & $    49$ & $^{+     1} _{-     2}$ & $       8$ & $            5$ & $^{+    7.1} _{-    3.9}$ & $    16$ & $^{+    21} _{-    11}$ & $       2$ & $            2$ & $^{+    2.0} _{-    1.0}$ & $     3$ & $^{+    17} _{-     3}$ & $       0$ & $            4$ & $^{+    0.4} _{-    0.2}$ & Rv3.1\\
 EB-FUV  73 & $    70$ & $^{+     1} _{-     1}$ & $       8$ & $            5$ & $^{+    3.0} _{-    2.2}$ & $    57$ & $^{+    12} _{-    11}$ & $       6$ & $            6$ & $^{+    2.5} _{-    1.8}$ & $    46$ & $^{+    24} _{-    22}$ & $       5$ & $            1$ & $^{+    2.1} _{-    1.5}$ & Rv3.1\\
 EB-FUV  74 & $   123$ & $^{+     1} _{-     2}$ & $          14$ & $ $ & $^{+    10.} _{-     6.}$ & $    84$ & $^{+    37} _{-    27}$ & $       9$ & $ 0$ & $^{+    6.8} _{-    3.9}$ & $    56$ & $^{+    65} _{-    38}$ & $       5$ & $            6$ & $^{+    4.6} _{-    2.5}$ & Rv3.1\\
 EB-FUV  75 & $    32$ & $^{+     1} _{-     6}$ & $          93$ & $ $ & $^{+   331.} _{-    42.}$ & $    13$ & $^{+    17} _{-    11}$ & $          30$ & $ $ & $^{+   120.} _{-    14.}$ & $     5$ & $^{+    25} _{-     5}$ & $       8$ & $            9$ & $^{+   40.2} _{-    4.4}$ & LMC2\\
 EB-FUV  76 & $   116$ & $^{+     1} _{-    14}$ & $          15$ & $ $ & $^{+    57.} _{-     5.}$ & $    84$ & $^{+    30} _{-    67}$ & $          10$ & $ $ & $^{+    45.} _{-     4.}$ & $    60$ & $^{+    54} _{-    60}$ & $       6$ & $            8$ & $^{+   33.7} _{-    2.7}$ & Rv3.1\\
 EB-FUV  77 & $     6$ & $^{+     1} _{-     1}$ & $       0$ & $            2$ & $^{+    1.1} _{-    0.0}$ & $     5$ & $^{+     0} _{-     5}$ & $       0$ & $            2$ & $^{+    1.2} _{-    0.0}$ & $     5$ & $^{+     0} _{-     5}$ & $       0$ & $            2$ & $^{+    1.3} _{-    0.0}$ & Rv3.1\\
\enddata
\tablenotetext{a}{Corresponding to the IDs in Table 1 and the blue labels in Fig.~\ref{fig:fuv_ha}.}
\tablenotetext{b}{For all sources, a foreground extinction of $\ebv=0.22$ was applied with MW-type dust, and  the 
 additional internal extinction  (from the total \ebv\ given in column 9 of Table 1) using three different dust types: 
`Rv3.1' stands for average MW extinction with Rv=3.1 \citep{Cardelli89}; `LMC' 
indicates the average LMC extinction for stars outside LMC2 by \citep{Misselt99}; `LMC2' indicates that the extinction curve by \citep{Misselt99} was used. }
\tablenotetext{c}{The results for the extinction curve given in this column were adopted in the analysis. 
only for source \# 6 we adopted an  SMC-type extinction curve, 
which gives 
age = $4^{+118}_{-4}$ Myr and mass = $11^{+43}_{-8}\times10^3$ \Msun}
\end{deluxetable}
\normalsize


\begin{thebibliography}


\bibitem[Bianchi et al.(2011)]{Bianchi11}  Bianchi, L., 2011, Ap\&SS, in press (special issue "UV Universe 2010", editors B. Shustov, A.I. Gomez de Castro, and M. Sachkov)
\bibitem[Bianchi et al.(2011)]{Bianchi10b}  Bianchi, L.,  et al. 2011, Ap\&SS, in press (special issue "UV Universe 2010", 
editors B. Shustov, A.I. Gomez de Castro, and M. Sachkov)
\bibitem[Bianchi et al.(2010)]{Bianchi10} Bianchi, L. , et al. 2010, AAS 215, 455.25 
\bibitem[Bianchi(2009)]{Bianchi09} Bianchi, L.\ 2009, Ap\&SS, 320, 11 
\bibitem[Bianchi(2007)]{Bianchi07a} Bianchi, L.\ 2007, in ``UV Astronomy: Stars from Birth to Death'', eds. A.I. Gomez de Castro and M. Barstow, UCM Editorial Complutense, p. 65
\bibitem[Bianchi \& Efremova (2006)]{BE06} Bianchi, L., \& Efremova, B.~V.\ 2006, \aj, 132, 378 
\bibitem[Bianchi et al.(2003a)]{Bianchi03a} Bianchi, L., et al. 2003a, Bulletin of the American Astronomical Society, 35, 1354 
\bibitem[Bianchi et al.(2003b)]{Bianchi03b} Bianchi, L., et al. 2003b, in ``The Local Group as an Astrophysical Laboratory'', STScI publication (M.Livio and T.Brown eds.), p. 10 
\bibitem[Bianchi et al.(2001)]{Bianchi01} Bianchi, L., Scuderi, S., Massey, P., \& Romaniello, M.\ 2001, \aj, 121, 2020  
\bibitem[Bianchi et al.(1999)]{Bianchi99}Bianchi, L., Chandar, R., Ford, H. 1999, Mem. SAIT, eds. D.de Martino \& L. Buson, Vol. 70 N.2, p. 629 
\bibitem[Boissier et al.(2007)]{Boissier07} Boissier, S., et al.\ 2007, \apjs, 173, 524 
\bibitem[Burgarella, D. et al. (2005)]{denis05}Burgarella, D., Buat, V.\& Iglesias-Páramo, J. 2005, \mnras, 360, 1413  
\bibitem[Calzetti(2001)]{Calzetti01} Calzetti, D.\ 2001, \pasp, 113, 1449 
\bibitem[Calzetti et al.(2005)]{Calzetti05} Calzetti, D., et al.\ 2005, \apj, 633, 871 
\bibitem[Cannon et al.(2006)]{Cannon06} Cannon, J.~M., et al.\ 2006, \apj, 652, 1170 
\bibitem[Cardelli et al.(1989)]{Cardelli89} Cardelli, J.~A., Clayton, G.~C., \& Mathis, J.~S.\ 1989, \apj, 345, 245 
\bibitem[Chandar et al. (1999)]{CBF99}Chandar, R., Bianchi, L., Ford, H. and Salasnich, B. 1999, PASP, 111, 794
\bibitem[Cortese et al.(2006)]{Cortese06} Cortese, L., et al.\ 2006, \apj, 637, 242 
\bibitem[de Blok \& Walter(2006)]{deBlok06} de Blok, W.~J.~G., \& Walter, F.\ 2006, \aj, 131, 343 
\bibitem[Gallart et al.(1996)]{Gallart96} Gallart, C., Aparicio, A., Bertelli, G., \& Chiosi, C.\ 1996, \aj, 112, 2596
\bibitem[Gil de Paz et al.(2007)]{dePaz07} Gil de Paz, A., et al.\ 2007, \apjs, 173, 185 
\bibitem[Girardi et al.(1995)]{Girardi95} Girardi, L., Chiosi, C., Bertelli, G., \& Bressan, A.\ 1995, \aap, 298, 87 
\bibitem[Efremov(2003)]{Efremov03} Efremov, Y.~N.\ 2003, Astron. Rep., 47, 1000 
\bibitem[Fall et al.(2009)]{Fall09} Fall, S.~M., Chandar, R., \& Whitmore, B.~C.\ 2009, \apj, 704, 453 
\bibitem[Fatuzzo \& Adams(2008)]{Fatuzzo08} Fatuzzo, M., \& Adams, F.~C.\ 2008, \apj, 675, 1361 
\bibitem[Gordon \& Clayton(1998)]
                   {Gordon98} Gordon, K.~D., \& Clayton, G.~C.\ 1998, \apj, 500, 816 
\bibitem[Hirashita et al.(2003)]{Hirashita03} Hirashita, H., Buat, V., \& Inoue, A.~K.\ 2003, \aap, 410, 83
\bibitem[Hodge(1977)]{Hodge77} Hodge, P.~W.\ 1977, \apjs, 33, 69 
\bibitem[Hodge et al.(1988)]{Hodge88} Hodge, P., Lee, M.~G., \& Kennicutt, R.~C., Jr.\ 1988, \pasp, 100, 917 
\bibitem[Hodge et al.(1989)]{Hodge89} Hodge, P., Lee, M.~G., \& Kennicutt, R.~C., Jr.\ 1989, \pasp, 101, 32 
\bibitem[Hubble(1925)]{Hubble25} Hubble, E.~P.\ 1925, \apj, 62, 409 
\bibitem[Hunter \& Elmegreen(2004)]{Hunter04} Hunter, D.~A., \& Elmegreen, B.~G.\ 2004, \aj, 128, 2170 
\bibitem[Hunter et al.(2010)]{Hunter10} Hunter, D.~A., Elmegreen, B.~G., \& Ludka, B.~C.\ 2010, \aj, 139, 447 
\bibitem[Kang et al.(2009)]{Kang09} Kang, Y., Bianchi, L., \& Rey, S.-C.\ 2009, \apj, 703, 614  
\bibitem[Karachentsev et al.(2004)]{Karachentsev04} 
                            Karachentsev, I.~D., Karachentseva, V.~E., Huchtmeier, W.~K., \& Makarov, D.~I.\ 2004, \aj, 127, 2031 
\bibitem[Kennicutt(1998)]{Kennicutt98} Kennicutt, R.~C., Jr.\ 1998, \araa, 36, 189
\bibitem[Killen \& Dufour(1982)]{Killen82} Killen, R.~M., \& Dufour, R.~J.\ 1982, \pasp, 94, 444
\bibitem[Kinman et al.(1979)]{Kinman79} Kinman, T.~D., Green, J.~R., \& Mahaffey, C.~T.\ 1979, \pasp, 91, 749 
\bibitem[Kroupa(2001)]{Kroupa01} Kroupa, P.\ 2001, \mnras, 322, 231 
\bibitem[Lada \& Lada(2003)]{Lada03} Lada, C.~J., \& Lada, E.~A.\ 2003, \araa, 41, 57 
\bibitem[Leroy et al.(2008)]{Leroy08} Leroy, A.~K., Walter, F., Brinks, E., Bigiel, F., de Blok, W.~J.~G., Madore, B., \& Thornley, M.~D.\ 2008, \aj, 136, 2782 
\bibitem[Martin et al.(2005)]{Martin05} Martin, D.~C., et al.\ 2005, \apjl, 619, L1 
\bibitem[Massey et al.(2007a)]{Massey07a} Massey, P., Olsen, K.~A.~G., Hodge, P.~W., Jacoby, G.~H., McNeill, R.~T., Smith, R.~C., \& Strong, S.~B.\ 2007a, \aj, 133, 2393
\bibitem[Massey et al.(2007b)]{Massey07b} Massey, P., McNeill, R.~T., Olsen, K.~A.~G., Hodge, P.~W., Blaha, C., Jacoby, G.~H., Smith, R.~C., \& Strong, S.~B.\ 2007b, \aj, 134, 2474 
\bibitem[McAlary et al.(1983)]{McAlary83} McAlary C.~W., Madore, B.~F., C.~W.,McGonegal, R., McLaren, R.~A., \& Welch, D.~L. \ 1983, \apj, 273, 539  
\bibitem[Melena et al.(2009)]{Melena09} Melena, N.~W., Elmegreen, B.~G., Hunter, D.~A., \& Zernow, L.\ 2009, \aj, 138, 1203 
\bibitem[Meurer et al.(2009)]{Meurer09} Meurer, G.~R., et al.\ 2009, \apj, 695, 765 
\bibitem[Misselt et al.(1999)]
           {Misselt99} Misselt, K.~A., Clayton, G.~C., \& Gordon, K.~D.\ 1999, \apj, 515, 128
\bibitem[Morrissey et al.(2007)]{Morrissey07} Morrissey, P., et al.\ 2007, \apjs, 173, 682 
\bibitem[Muschielok et al.(1999)]{Muschielok99} Muschielok, B., et al.\ 1999, \aap, 352, L40
\bibitem[O'Dell et al.(1999)]{ODell99} O'Dell, C.~R., Hodge, P.~W., \& Kennicutt, R.~C., Jr.\ 1999, \pasp, 111, 1382 
\bibitem[Pagel et al.(1980)]{Pagel80} Pagel, B.~E.~J., Edmunds, M.~G., \& Smith, G.\ 1980, \mnras, 193, 219
\bibitem[Panuzzo et al. (2003)]{Panuzzo03} Panuzzo, P. Bressan, A., Silva, L. et al. 2003, \aap , 409, 99
\bibitem[Pietrzy{\'n}ski et al.(2004)]{Pietrzynski04} Pietrzy{\'n}ski, G., Gieren, W., Udalski, A., Bresolin, F., Kudritzki, R.-P., Soszy{\'n}ski, I., Szyma{\'n}ski, M., \& Kubiak, M.\ 2004, \aj, 128, 2815
\bibitem[Venn et al.(2001)]{Venn01} Venn, K.~A., et al.\ 2001, \apj, 547, 765 
\bibitem[Wyder et al.(2007)]{Wyder07} Wyder, T.~K., et al.\ 2007, \apjs, 173, 293 

\end{thebibliography}
\end{document}